\documentclass[12pt]{article}
\usepackage{amssymb,amscd,array}
\catcode `\@=11
\@addtoreset{equation}{section}

\newtheorem{thm}{Theorem}[section]

\newtheorem{lemma}[thm]{Lemma}

\def\qed{\blacksquare}
\newcommand{\be}{\begin{equation}}
\newcommand{\ee}{\end{equation}}
\newcommand{\bea}{\begin{eqnarray}}
\newcommand{\eea}{\end{eqnarray}}
\newcommand{\R}{\mathbb{R}}
\newcommand{\N}{\mathbb{N}}
\newcommand{\C}{\mathbb{C}}

\begin{document}
\begin{titlepage}

\begin{center}
{\bf \Large{Loop Anomalies in the Causal Approach  \\}}
\end{center}
\vskip 1.0truecm
\centerline{D. R. Grigore, 
\footnote{e-mail: grigore@theory.nipne.ro}}
\vskip5mm
\centerline{Department of Theoretical Physics,}
\centerline{Institute for Physics and Nuclear Engineering ``Horia Hulubei"}
\centerline{Bucharest-M\u agurele, P. O. Box MG 6, ROM\^ANIA}

\vskip 2cm
\bigskip \nopagebreak
\begin{abstract}
\noindent
We consider gauge models in the causal approach and study one-loop
contributions to the chronological products and the anomalies they produce. 
We prove that in order greater than 4 there are no one-loop anomalies.
Next we analyze one-loop anomalies in the second and third order of the perturbation theory. We prove that the
even parity contributions (with respect to parity) do not produce anomalies; for the odd parity
contributions we reobtain the well-known result.

\end{abstract}
\end{titlepage}

\section{Introduction}

The general framework of perturbation theory consists in the construction of 
the chronological products such that Bogoliubov axioms are verified \cite{BS},
\cite{EG}, \cite{DF}; for every set of Wick monomials 
$ 
W_{1}(x_{1}),\dots,W_{n}(x_{n}) 
$
acting in some Fock space
$
{\cal H}
$
one associates the operator-valued distributions
$ 
T^{W_{1},\dots,W_{n}}(x_{1},\dots,x_{n})
$  
called chronological products; it will be convenient to use another notation: 
$ 
T(W_{1}(x_{1}),\dots,W_{n}(x_{n})). 
$ 
The construction of the chronological products can be done recursively according
to Epstein-Glaser prescription \cite{EG}, \cite{Gl} (which reduces the induction
procedure to a distribution splitting of some distributions with causal support)
or according to Stora prescription \cite{PS} (which reduces the renormalization
procedure to the process of extension of distributions). These products are not
uniquely defined but there are some natural limitation on the arbitrariness. If
the arbitrariness does not grow with $n$ we have a renormalizable theory. An
equivalent point of view uses retarded products \cite{St1}.

Gauge theories describe particles of higher spin. Usually such theories are not
renormalizable. However, one can save renormalizablility using ghost fields.
Such theories are defined in a Fock space
$
{\cal H}
$
with indefinite metric, generated by physical and un-physical fields (called
{\it ghost fields}). One selects the physical states assuming the existence of
an operator $Q$ called {\it gauge charge} which verifies
$
Q^{2} = 0
$
and such that the {\it physical Hilbert space} is by definition
$
{\cal H}_{\rm phys} \equiv Ker(Q)/Im(Q).
$
The space
$
{\cal H}
$
is endowed with a grading (usually called {\it ghost number}) and by
construction the gauge charge is raising the ghost number of a state. Moreover,
the space of Wick monomials in
$
{\cal H}
$
is also endowed with a grading which follows by assigning a ghost number to
every one of the free fields generating
$
{\cal H}.
$
The graded commutator
$
d_{Q}
$
of the gauge charge with any operator $A$ of fixed ghost number
\be
d_{Q}A = [Q,A]
\ee
is raising the ghost number by a unit. It means that
$
d_{Q}
$
is a co-chain operator in the space of Wick polynomials. From now on
$
[\cdot,\cdot]
$
denotes the graded commutator.
 
A gauge theory assumes also that there exists a Wick polynomial of null ghost
number
$
T(x)
$
called {\it the interaction Lagrangian} such that
\be
~[Q, T] = i \partial_{\mu}T^{\mu}
\label{gau1}
\ee
for some other Wick polynomials
$
T^{\mu}.
$
This relation means that the expression $T$ leaves invariant the physical
states, at least in the adiabatic limit. Indeed, if this is true we have:
\be
T(f)~{\cal H}_{\rm phys}~\subset~~{\cal H}_{\rm phys}  
\label{gau2}
\ee
up to terms which can be made as small as desired (making the test function $f$
flatter and flatter). In all known models one finds out that there exists a
chain of Wick polynomials
$
T^{\mu},~T^{\mu\nu},~T^{\mu\nu\rho},\dots
$
such that:
\be
~[Q, T] = i \partial_{\mu}T^{\mu}, \quad
[Q, T^{\mu}] = i \partial_{\nu}T^{\mu\nu}, \quad
[Q, T^{\mu\nu}] = i \partial_{\rho}T^{\mu\nu\rho},\dots
\label{descent}
\ee
It so happens that for all these models the expressions
$
T^{\mu\nu},~T^{\mu\nu\rho},\dots
$
are completely antisymmetric in all indexes; it follows that the chain of
relation stops at the step $4$ (if we work in four dimensions). We can also use
a compact notation
$
T^{I}
$
where $I$ is a collection of induces
$
I = [\nu_{1},\dots,\nu_{p}]~(p = 0,1,\dots,)
$
and the brackets emphasize the complete antisymmetry in these induces. All these
polynomials have the same canonical dimension
\be
\omega(T^{I}) = \omega_{0},~\forall I
\ee
and because the ghost number of
$
T \equiv T^{\emptyset}
$
is supposed null, then we also have:
\be
gh(T^{I}) = |I|.
\ee
One can write compactly the relations (\ref{descent}) as follows:
\be
d_{Q}T^{I} = i~\partial_{\mu}T^{I\mu}.
\label{descent1}
\ee

For concrete models the equations (\ref{descent}) can stop earlier: for 
instance in the Yang-Mills case we have
$
T^{\mu\nu\rho} = 0
$
and in the case of gravity
$
T^{\mu\nu\rho\sigma} = 0.
$

Now we can construct the chronological products
\be
T^{I_{1},\dots,I_{n}}(x_{1},\dots,x_{n}) \equiv
T(T^{I_{1}}(x_{1}),\dots,T^{I_{n}}(x_{n}))
\label{chronos}
\ee
according to the recursive procedure. We say that the theory is gauge invariant
in all orders of the perturbation theory if the following set of identities
generalizing (\ref{descent1}):
\be
d_{Q}T^{I_{1},\dots,I_{n}} = 
i \sum_{l=1}^{n} (-1)^{s_{l}} {\partial\over \partial x^{\mu}_{l}}
T^{I_{1},\dots,I_{l}\mu,\dots,I_{n}}
\label{gauge}
\ee
are true for all 
$n \in \N$
and all
$
I_{1}, \dots, I_{n}.
$
Here we have defined
\be
s_{l} \equiv \sum_{j=1}^{l-1} |I|_{j}.
\ee
In particular, the case
$
I_{1} = \dots = I_{n} = \emptyset
$
it is sufficient for the gauge invariance of the scattering matrix, at least
in the adiabatic limit: we have the same argument as for relation (\ref{gau2}).

Such identities can be usually broken by {\it anomalies} i.e. expressions of the
type
$
A^{I_{1},\dots,I_{n}}
$
which are quasi-local and might appear in the right-hand side of the relation
(\ref{gauge}). In a previous paper we have emphasized the cohomological
structure of this problem \cite{sr3}. We consider a {\it cochain} to be
an ensemble of distribution-valued operators of the form
$
C^{I_{1},\dots,I_{n}}(x_{1},\dots,x_{n}),~n = 1,2,\cdots
$
(usually we impose some supplementary symmetry properties) and define the
derivative operator $\delta$ according to
\be
(\delta C)^{I_{1},\dots,I_{n}}
= \sum_{l=1}^{n} (-1)^{s_{l}} {\partial\over \partial x^{\mu}_{l}}
C^{I_{1},\dots,I_{l}\mu,\dots,I_{n}}.
\ee
We can prove that 
\be
\delta^{2} = 0.
\ee
Next we define
\be
s = d_{Q} - i \delta,\qquad \bar{s} = d_{Q} + i \delta
\ee
and note that
\be
s \bar{s} = \bar{s} s = 0.
\ee
We call {\it relative cocycles} the expressions $C$ verifying
\be
sC = 0
\ee
and a {\it relative coboundary} an expression $C$ of the form
\be
C = \bar{s}B.
\ee
The relation (\ref{gauge}) is simply the cocycle condition
\be
sT = 0.
\ee

If we can prove that this relation is valid up to the order 
$
n - 1
$
then in order $n$ this relation is valid up to {\it anomalies}:
\be
sT = {\cal A}
\ee
where the anomalies in the right hand side have the generic form
\be
{\cal A}(x_{1},\dots,x_{n}) = \sum p_{i}(\partial)\delta(x_{1},\dots,x_{n})~
W_{i}(x_{1},\dots,x_{n}).
\ee
Here 
\be
\delta(x_{1},\dots,x_{n}) 
= \delta(x_{1} - x_{n}) \cdots \delta(x_{n-1} - x_{n}),
\ee
$
p_{i}
$
are polynomials in the partial derivatives and 
$
W_{i}
$
are Wick polynomials. There is a bound on the number
\be
deg({\cal A}) \equiv supp_{i}~\{deg(p_{i}) + \omega(W_{i})\}
\ee
coming from the power counting theorem; here 
$
deg(p)
$
is the degree of the polynomial $p$ and
$
\omega(W)
$
is the canonical dimension of the Wick polynomial $W$. We call this number
the {\it canonical dimension of the anomaly}. For instance if the interaction
Lagrangian and the associated expressions
$
T^{I}
$
verify
$
\omega(T^{I}) = 4
$
(as is the case of Yang-Mills models) then the canonical dimension of the
anomaly is 
$
\leq 5
$. 
The contributions corresponding to maximal degree will be called {\it dominant}.

Gauge theories have been intensively studied in another formalism based on
functional integrations and Green functions. There is no proof of the
equivalence between the functional formalism and the causal formalism which we
use here. A supplementary problem in the functional formalism is that the Green
functions are affected by infra-red divergences; an adiabatic limit must be
performed and, as it can be seen from the paper of Epstein and Glaser, this
limit is not easy to perform. 

So, for the moment, it is safer to consider the causal formalism is not
equivalent to the functional formalism and study gauge theories in an
independent way. In particular, the problem of anomalies produced by loop
contributions is very interesting. No systematic study is available for the loop
contributions in the third order of the perturbation theory in the causal
approach. We propose to do this in this paper. The basic idea is
to isolate some typical numerical distributions with causal support appearing
in the loop contributions in the second and the third order of the
perturbation theory; then we prove that some identities verified by these
distributions can be causally split without anomalies. This idea is in the
spirit of the master Ward identity considered in the literature \cite{DB},
\cite{DF1}, but the actual proof of our identities seems to be considerably
different. 

In the next Section we will give a minimal account of the gauge
theories in the causal approach. Then in Section \ref{one-loop} we make a general
analysis of the one-loop contributions in arbitrary order of the perturbation
theory. As a result we prove that for 
$
N > 4
$
there are no one-loop anomalies. So, next we turn to the one-loop anomalies in the second 
and third order of perturbation theory in Sections \ref{second} and \ref{third}.

\newpage

\section{General Gauge Theories\label{ggt}}
\subsection{Perturbation Theory}
 
We give here the essential ingredients of perturbation theory. Suppose that the
Wick monomials
$
W_{1},\dots,W_{n}
$
are self-adjoint:
$
W_{j}^{\dagger} = W_{j},~\forall j = 1,\dots,n.
$
The chronological products
$ 
T(W_{1}(x_{1}),\dots,W_{n}(x_{n})) \quad n = 1,2,\dots
$
are verifying the following set of axioms:
\begin{itemize}
\item
Skew-symmetry in all arguments
$
W_{1}(x_{1}),\dots,W_{n}(x_{n}):
$
\be
T(\dots,W_{i}(x_{i}),W_{i+1}(x_{i+1}),\dots,) =
(-1)^{f_{i} f_{i+1}} T(\dots,W_{i+1}(x_{i+1}),W_{i}(x_{i}),\dots)
\ee
where
$f_{i}$
is the number of Fermi fields appearing in the Wick monomial
$W_{i}$.
\item
Poincar\'e invariance: we have a natural action of the Poincar\'e group in the
space of Wick monomials and we impose that for all 
$(a,A) \in inSL(2,\C)$
we have:
\be
U_{a, A} T(W_{1}(x_{1}),\dots,W_{n}(x_{n})) U^{-1}_{a, A} =
T(A\cdot W_{1}(A\cdot x_{1}+a),\dots,A\cdot W_{n}(A\cdot x_{n}+a));
\label{invariance}
\ee

Sometimes it is possible to supplement this axiom by other invariance
properties: space and/or time inversion, charge conjugation invariance, global
symmetry invariance with respect to some internal symmetry group, supersymmetry,
etc.
\item
Causality: if
$x_{i} \geq x_{j}, \quad \forall i \leq k, \quad j \geq k+1$
then we have:
\be
T(W_{1}(x_{1}),\dots,W_{n}(x_{n})) =
T(W_{1}(x_{1}),\dots,W_{k}(x_{k}))~~T(W_{k+1}(x_{k+1}),\dots,W_{n}(x_{n}));
\label{causality}
\ee
\item
Unitarity: We define the {\it anti-chronological products} according to
\be
(-1)^{n} \bar{T}(W_{1}(x_{1}),\dots,W_{n}(x_{n})) \equiv \sum_{r=1}^{n} 
(-1)^{r} \sum_{I_{1},\dots,I_{r} \in Part(\{1,\dots,n\})}
\epsilon~~T_{I_{1}}(X_{1})\cdots T_{I_{r}}(X_{r})
\label{antichrono}
\ee
where the we have used the notation:
\be
T_{\{i_{1},\dots,i_{k}\}}(x_{i_{1}},\dots,x_{i_{k}}) \equiv 
T(W_{i_{1}}(x_{i_{1}}),\dots,W_{i_{k}}(x_{i_{k}}))
\ee
and the sign
$\epsilon$
counts the permutations of the Fermi factors. Then the unitarity axiom is:
\be
\bar{T}(W_{1}(x_{1}),\dots,W_{n}(x_{n})) =
T(W_{1}(x_{1}),\dots,W_{n}(x_{n}))^{\dagger}.
\label{unitarity}
\ee
\item
The ``initial condition"
\be
T(W(x)) = W(x).
\ee
\end{itemize}

It can be proved that this system of axioms can be supplemented with
\bea
T(W_{1}(x_{1}),\dots,W_{n}(x_{n}))
\nonumber \\
= \sum \quad
<\Omega, T(W^{\prime}_{1}(x_{1}),\dots,W^{\prime}_{n}(x_{n}))\Omega>~~
:W^{\prime\prime}_{1}(x_{1}),\dots,W^{\prime\prime}_{n}(x_{n}):
\label{wick-chrono2}
\eea
where
$W^{\prime}_{i}$
and
$W^{\prime\prime}_{i}$
are Wick submonomials of
$W_{i}$
such that
$W_{i} = :W^{\prime}_{i} W^{\prime\prime}_{i}:$
and appropriate signs should be included if Fermi fields are present; here
$\Omega$
is the vacuum state. This is called the {\it Wick expansion property}. 

We can also include in the induction hypothesis a limitation on the order of
singularity of the vacuum averages of the chronological products associated to
arbitrary Wick monomials
$W_{1},\dots,W_{n}$;
explicitly:
\be
\omega(<\Omega, T^{W_{1},\dots,W_{n}}(X)\Omega>) \leq
\sum_{l=1}^{n} \omega(W_{l}) - 4(n-1)
\label{power}
\ee
where by
$\omega(d)$
we mean the order of singularity of the (numerical) distribution $d$ and by
$\omega(W)$
we mean the canonical dimension of the Wick monomial $W$; in particular this
means
that we have
\be
T(W_{1}(x_{1}),\dots,W_{n}(x_{n}))
= \sum_{g} t_{g}(x_{1},\dots,x_{n})~W_{g}(x_{1},\dots,x_{n})
\label{generic}
\ee
where
$W_{g}$
are Wick polynomials of fixed canonical dimension and
$t_{g}$
are distributions in 
$
n - 1
$
variables (because of translation invariance) with the order of singularity
bounded by the power counting
theorem \cite{EG}:
\be
\omega(t_{g}) + \omega(W_{g}) \leq
\sum_{j=1}^{n} \omega(W_{j}) - 4 (n - 1)
\label{power1}
\ee
and the sum over $g$ is essentially a sum over Feynman graphs. The
contributions verifying the strict inequality above i.e. with the strict
inequality 
$<$
sign, will be called {\it super-renormalizable} as in \cite{sr3}. The contributions
saturating the inequality (i.e. corresponding to the equal sign) will
 be called {\it dominant}; they will produce dominant anomalies.

Up to now, we have defined the chronological products only for self-adjoint Wick
monomials 
$
W_{1},\dots,W_{n}
$
but we can extend the definition for arbitrary  Wick polynomials by linearity.

One can modify the chronological products without destroying the basic property
of causality {\it iff} one can make
\be
t_{g} \rightarrow t_{g} + P_{g}(\partial)
\delta(x_{1} - x_{n})\cdots\delta(x_{n-1} - x_{n})
\label{renorm}
\ee
with 
$P_{g}$ 
a monomials in the partial derivatives. If we want to preserve (\ref{power1}) we
impose the restriction
\be
deg(P_{g}) + \omega(W_{g}) \leq
\sum_{j=1}^{n} \omega(W_{j}) - 4 (n - 1)
\label{power2}
\ee
and some other restrictions are following from Lorentz covariance and unitarity.

From now on we consider that we work in the four-dimensional Minkowski space and
we have the Wick polynomials
$
T^{I}
$
such that the descent equations (\ref{descent1}) are true and we also have
\be
T^{I}(x_{1})~T^{J}(x_{2}) = (-1)^{|I||J|}~T^{J}(x_{2})~T^{I}(x_{1}),~~
\forall~x_{1} \sim x_{2}
\label{graded-comm}
\ee
i.e. for 
$
x_{1} - x_{2}
$
space-like these expressions causally commute in the graded sense. The
chronological products 
$
T^{I_{1},\dots,I_{n}}(x_{1},\dots,x_{n})
$
are constructed according to the prescription (\ref{chronos}) from the
Introduction and they form a cohomological object.
One way to obtain them is to proceed recursively. For instance, we can define
the {\it causal commutator} according to:
\be
D^{IJ}(x_{1},x_{2}) = T^{I}(x_{1})~T^{J}(x_{2}) - (-1)^{|I||J|}~T^{J}(x_{2})~T^{I}(x_{1})
\ee
and after the operation of causal splitting one can obtain the second order 
chronological products. Generalizations of this formula are available for higher 
orders of the perturbation theory.

\subsection{Gauge Theories}
We will be interested in the following by Yang-Mills models. The Hilbert space of the model is generated by
the following types of particles:

1. Particles of null mass and helicity $1$ (photons and gluons). They are described by the vector fields 
$
v^{\mu}_{a}
$ 
(with Bose statistics) and the scalar fields 
$
u_{a}, \tilde{u}_{a}
$
(with Fermi statistics) where 
$
a \in I_{1}
$
with 
$
I_{1}
$
and index set of cardinal 
$
r_{1};
$
all these fields have null mass.

2. Particles of positive mass and spin $1$ (heavy Bosons). They are described by the vector fields 
$
v^{\mu}_{a}
$ 
(with Bose statistics) and the scalar fields 
$
u_{a}, \tilde{u}_{a}
$
(with Fermi statistics) and scalar fields
$
\Phi_{a}
$ 
where 
$
a \in I_{2}
$
with 
$
I_{2}
$
and index set of cardinal 
$
r_{2};
$
all these fields have mass
$
m_{a}.
$

3. Scalar particles (essentially we have only the Higgs particle but we consider more for generality) 
$
\Phi_{a}
$ 
where 
$
a \in I_{3}
$
with 
$
I_{3}
$
and index set of cardinal 
$
r_{3};
$
these fields have mass
$
m_{a}^{H}.
$

4. Dirac fields 
$
\psi_{A}
$
where 
$
A \in I_{4}
$
with 
$
I_{4}
$
and index set of cardinal 
$
r_{4};
$
these fields have mass
$
M_{A}.
$

To describe completely the model we need to give the following elements:

- The $2$-point functions; then we can generate the $n$-point functions using as a guide Wick theorem.

- A Hermiticity structure.

- The action of the gauge charge on the fields.

All these elements can be found in preceding publications for instance \cite{cohomology}. One can use
the formalism described there to obtain in an unique way the expression of the interaction Lagrangian
$T$: it is (relatively) cohomologous to a non-trivial co-cycle of the form:
\bea
T = f_{abc} \left( {1\over 2}~v_{a\mu}~v_{b\nu}~F_{c}^{\nu\mu}
+ u_{a}~v_{b}^{\mu}~\partial_{\mu}\tilde{u}_{c}\right)
\nonumber \\
+ f^{\prime}_{abc} (\Phi_{a}~\phi_{b}^{\mu}~v_{c\mu} +
m_{b}~\Phi_{a}~\tilde{u}_{b}~u_{c})
\nonumber \\
+ {1\over 3!}~f^{\prime\prime}_{abc}~\Phi_{a}~\Phi_{b}~\Phi_{c}
+ j^{\mu}_{a}~v_{a\mu} + j_{a}~\Phi_{a};
\eea
where there are various relations between the constants appearing above. The first line give the pure Yang-Mills 
interaction, the second line is the vector-scalar interaction, then comes the pure scalar interaction and the last two
terms give the interaction of the Dirac fields with the vector and resp. scalar particles mediated by the vector and scalar currents
\bea
j_{a}^{\mu} = \sum_{\epsilon}~
\overline{\psi} t^{\epsilon}_{a} \otimes \gamma^{\mu}\gamma_{\epsilon} \psi
\qquad
j_{a} = \sum_{\epsilon}~
\overline{\psi} s^{\epsilon}_{a} \otimes \gamma_{\epsilon} \psi
\label{current}
\eea
The expression above is constrained by Lorentz invariance and the bound $< 4$ on the canonical dimension. 
One can give explicit formulas for the associated expressions
$
T^{\mu}, T^{\mu\nu}
$
(see the ref. cited above).

\subsection{Distributions with Causal Support and Causal Splitting}

We will use many times the so-called {\it central splitting of
causal distributions} \cite{Sc2}. We remind the reader the basic formula. Let 
$
d \in ({\cal S}^{4n})^{\prime}
$
be a distribution in the variables
$
x_{1},\dots,x_{n}
$
from the Minkowski space. Suppose that $d$ has causal support i.e.
\be
supp(d) \in \{(x_{1},\dots,x_{n}) | x_{j} - x_{n} \in V^{+} \cup V^{-}, j =
1,\dots,n - 1\}
\ee
and has the order of causality 
$
\omega = \omega(d) \in \N;
$
essentially this means that the Fourier transform 
$
\tilde{d}
$
of $d$ behaves for large momenta as
$
p^{\omega}.
$
It is a standard theorem in distribution theory that we can split
\be
d = a - r
\ee
where
\bea
supp(a) \in \{(x_{1},\dots,x_{n}) | x_{j} - x_{n} \in V^{+}, j = 1,\dots,n - 1\}
\nonumber \\
supp(r) \in \{(x_{1},\dots,x_{n}) | x_{j} - x_{n} \in V^{-}, j = 1,\dots,n - 1\}
\eea
are called the {\it advanced} and resp. {\it retarded} components of $d$. If
$
\omega(d) < 0
$
then $a$ and $r$ are uniquely determined; formally we have 
\bea
a(x) = \theta^{+}(x)~d(x)
\nonumber \\
r(x) = \theta^{-}(x)~d(x)
\eea
where
$
\theta^{\pm}
$
are some Heaviside functions separating the two pieces of the light cones. Let
us suppose that 
$
0 \not\in supp(\tilde{d});
$
then taking the Fourier transform we get for:
\be
\tilde{a}(p) = {i \over 2\pi} \int_{-\infty}^{\infty} dt 
{\tilde{d}(t p) \over 1 - t + i0},\qquad p \in V^{+} \cup V^{-}
\label{central1}
\ee
and the integral is convergent. If 
$
\omega(d) \geq 0
$
then the integral is not convergent any more and (as for the subtracted Cauchy
formula) we have:
\be
\tilde{a}(p) = {i \over 2\pi} \int_{-\infty}^{\infty} dt 
{\tilde{d}(t p) \over (t - i0)^{\omega}~(1 - t + i0)}
\label{central2}
\ee
and the integral is again convergent.
\newpage

\section{One Loop Feynman Distributions\label{one-loop}}

We consider here one-loop contributions in arbitrary order $n$ of the perturbation theory. 
We consider the Feynman distribution 
$
D^{F}_{m} \in {\cal S}^{\prime}(\R^{4})
$
for mass
$
m \geq 0;
$
it is known that this distribution has the order of singularity
$
\omega = - 2.
$

We now define some Feynman distribution from
$
{\cal S}^{\prime}(\R^{4n}).
$
We define the diagonal domain
\be
D_{n} \equiv \{(x_{1},\dots,x_{n}) \in \R^{4n} \times \cdots \times\R^{4n} |
x_{1} = \cdots = x_{n} \}
\ee
and note that for
$
(x_{1},\cdots,x_{n}) \not\in D_{n}
$
the expression
\be
d_{m_{1},\cdots,m_{n}}^{(0)}(x_{1},\dots,x_{n}) \equiv
D^{F}_{m_{1}}(x_{1} - x_{n})~D_{m_{2}}^{F}(x_{2} - x_{3})~\dots D_{m_{n}}^{F}(x_{n} - x_{1})
\ee
is well defined and has the order of singularity
\be
\omega(d_{m_{1},\cdots,m_{n}}^{(0)}) = 4 - 2n.
\ee
The same goes true for the associated distributions
\be
{\cal D}_{j}^{\mu}d_{m_{1},\cdots,m_{n}}^{(0)} \equiv
D^{F}_{m_{1}}(x_{1} - x_{n})~\dots \partial^{\mu}D_{m_{j}}^{F}(x_{l} - x_{l+1})~\dots D_{m_{n}}^{F}(x_{n} - x_{1})
\ee
etc. and the order of singularity increases with one unit for every ``derivative"
$
{\cal D}.
$
But according to some standard theorems, these distributions can be
extended to 
$
D_{n}
$
also in such a way that
the order of singularity, translation invariance and Lorentz covariance are
preserved. We denote these distributions by
$
d_{m_{1},\cdots,m_{n}}^{F},~
({\cal D}_{l}^{\mu}d_{m_{1},\cdots,m_{n}})^{F},
$
etc. First we have an elementary result
\begin{thm}
The following formulas are true
\be
{\partial \over \partial x_{j\mu}}d_{m_{1},\cdots,m_{n}}^{(0)} = 
{\cal D}_{j}^{\mu}d_{m_{1},\cdots,m_{n}}^{(0)} 
- {\cal D}_{j-1}^{\mu}d_{m_{1},\cdots,m_{n}}^{(0)},~j = 1,\dots,n
\label{d1}
\ee
where we convene that
$
{\cal D}_{0}^{\mu} \equiv {\cal D}_{n}^{\mu}.
$
These relations remain true for the corresponding Feynman propagators:
\be
{\partial \over \partial x_{j\mu}}d_{m_{1},\cdots,m_{n}}^{F} = 
({\cal D}_{j}^{\mu}d_{m_{1},\cdots,m_{n}})^{F} 
- ({\cal D}_{j-1}^{\mu}d_{m_{1},\cdots,m_{n}})^{F},~j = 1,\dots,n
\label{d1-F}
\ee
\end{thm}
{\bf Proof:} The proof of formula (\ref{d1}) is elementary. When we
extend the formula to the diagonal set 
$
D_{n}
$
we use standard results in distribution theory and get
\be
{\partial \over \partial x_{j\mu}}d_{m_{1},\cdots,m_{n}}^{F} = 
({\cal D}_{j}^{\mu}d_{m_{1},\cdots,m_{n}})^{F} 
- ({\cal D}_{j-1}^{\mu}d_{m_{1},\cdots,m_{n}})^{F}
+ p^{\mu}_{j}(\partial_{1},\dots,\partial_{n-1})\delta(x_{1},\dots,x_{n}),~j = 1,\dots,n
\label{d1-e}
\ee
where the last expression is an ``anomaly" i.e. an expression with support on the diagonal set 
$
D_{n}:
$
this means that
$
p^{\mu}_{j}
$
is a polynomial in the partial derivatives and 
\be
\delta(x_{1},\dots,x_{n}) \equiv \delta(x_{1} - x_{n})\cdots \delta(x_{n-1} - x_{n})
\ee
is the delta distribution associated to the diagonal set 
$
D_{n}.
$

Because the extensions can be done preserving the order of singularity we obtain
$
deg(p^{\mu}_{j}) < 0
$
so in fact
$
p^{\mu}_{j} = 0
$
i.e. there are no anomalies.
$\qed$

We can do this proof in a different way. First we ``solve" the relations
(\ref{d1}):
\begin{lemma}
The following formulas are true:
\be
{\cal D}_{j}^{\mu}d^{(0)} = {\cal D}_{n}^{\mu}d^{(0)} 
+ \sum_{p=1}^{j} {\partial \over \partial x_{p\mu}}d^{(0)},~
j = 1,\dots,n.
\label{d1-s}
\ee
\end{lemma}
The proof is elementary: we have to check that (\ref{d1-s}) verifies
identically (\ref{d1}). Also we notice that the relation (\ref{d1-s})
is consistent: if we take 
$
l = n
$
then we obtain
\be
\sum_{p=1}^{n} {\partial \over \partial x_{p\mu}}d^{(0)} = 0
\ee
which is the infinitesimal form of the translation invariance of 
$
d^{(0)}
$. 
Now we consider a convenient choice for 
$
d^{F}
$
and
$
{\cal D}_{n}^{\mu}d^{F}
$
and {\bf define}
\be
{\cal D}_{j}^{\mu}d^{F} \equiv {\cal D}_{n}^{\mu}d^{F} 
+ \sum_{p=1}^{j} {\partial \over \partial x_{p\mu}}d^{F},~
j = 1,\dots,n.
\label{d1-sF}
\ee
It follows that we have (\ref{d1-e}) and the translation invariance
\be
\sum_{p=1}^{n} {\partial \over \partial x_{p\mu}}d^{F} = 0
\ee
of the extended distribution
$
d^{F}.
$
So we can obtain the formulas (\ref{d1-e}) considering convenient choices for 
$
d^{F}
$
and
$
{\cal D}_{n}^{\mu}d^{F}
$
and then {\bf defining} 
$
{\cal D}_{l}^{\mu}d^{F}
$
for
$
l = 1,\dots, n - 1
$
through (\ref{d1-sF}).

We proceed in the same way for similar identities.
\begin{thm}
The following formulas are true:
\be
{\partial \over \partial x_{j\mu}}{\cal D}^{\nu}_{k}d^{(0)} = 
{\cal D}_{j}^{\mu}{\cal D}^{\nu}_{k}d^{(0)} 
- {\cal D}_{j - 1}^{\mu}{\cal D}^{\nu}_{k}d^{(0)}
\label{d2}
\ee
\end{thm}
The proof is elementary. Now we have a generalization of formula (\ref{d1-s}):
\begin{thm}
The following formulas are true for all
$
j, k = 1,\dots,n:
$
\be
{\cal D}_{j}^{\mu}{\cal D}^{\nu}_{k}d^{(0)} 
= {\cal D}_{n}^{\mu}{\cal D}^{\nu}_{n}d^{(0)}
+ \sum_{p=1}^{j} {\partial \over \partial x_{p\mu}}{\cal D}^{\nu}_{n}d^{(0)} 
+ \sum_{q=1}^{k} {\partial \over \partial x_{q\nu}}{\cal D}^{\mu}_{n}d^{(0)} 
+ \sum_{p=1}^{j} \sum_{q=1}^{k} {\partial^{2} \over \partial x_{p\mu}\partial x_{q\nu}}d^{(0)}. 
\ee
\label{d2-s}
\end{thm}
We first note that the preceding formula is consistent: we have symmetry for
$
l \leftrightarrow k,~~ \mu \leftrightarrow \nu
$
and if we take $j$ and /or $k$ equal to $n$ we get an identity due to
the translation invariance property. It is elementary to prove that the 
preceding formula verifies identically (\ref{d2}). As before we have:
\begin{thm}
We can extend the distributions is such a way that we have
\be
{\partial \over \partial x_{j\mu}}{\cal D}^{\nu}_{k}d^{F} = 
({\cal D}_{j}^{\mu}{\cal D}^{\nu}_{k}d)^{F} 
- ({\cal D}_{j - 1}^{\mu}{\cal D}^{\nu}_{k}d)^{F}
\label{d2-F}
\ee
\end{thm}
{\bf Proof:}  We start from some convenient choice for
$
d^{F}, ({\cal D}_{n}^{\mu}d)^{F}, ({\cal D}_{n}^{\mu}{\cal D}^{\nu}_{n}d)^{F}
$
and define the other distributions
$
({\cal D}_{l}^{\mu}{\cal D}^{\nu}_{k}d)^{F}
$
by relations obtained by the previous ones modified with an appropriate upper
index F:
\be
({\cal D}_{j}^{\mu}{\cal D}^{\nu}_{k}d)^{F} 
\equiv ({\cal D}_{n}^{\mu}{\cal D}^{\nu}_{n}d)^{F}
+ \sum_{p=1}^{j} {\partial \over \partial x_{p\mu}}({\cal D}^{\nu}_{n}d)^{F} 
+ \sum_{q=1}^{k} {\partial \over \partial x_{q\nu}}({\cal D}^{\mu}_{n}d)^{F} 
+ \sum_{p=1}^{j} \sum_{q=1}^{k} {\partial^{2} \over \partial x_{p\mu}\partial x_{q\nu}}d^{F}. 
\ee
\label{d2-sF}
The definitions are consistent and we have (\ref{d2-F}) from the statement.
$\qed$

A generalization of the preceding formulas is available. 
\begin{thm}
The following formulas are true:
\be
{\partial \over \partial x_{j\mu}}
{\cal D}^{\nu_{1}}_{k_{1}}\dots {\cal D}^{\nu_{p}}_{k_{p}}d^{(0)} = 
{\cal D}_{j}^{\mu}{\cal D}^{\nu_{1}}_{k_{1}}\dots {\cal D}^{\nu_{p}}_{k_{p}}d^{(0)} 
- {\cal D}_{j - 1}^{\mu}{\cal D}^{\nu_{1}}_{k_{1}}\dots {\cal D}^{\nu_{p}}_{k_{p}}d^{(0)}
\label{dp}
\ee
\end{thm}
Next we define the operators
\be
d_{k}^{\mu} \equiv \sum_{l=1}^{k} {\partial \over \partial x_{l\mu}}
\ee
and obtain the generalization of formula (\ref{d2-s}):
\begin{thm}
The following formulas are true:
\be
{\cal D}^{\nu_{1}}_{k_{1}}\dots {\cal D}^{\nu_{p}}_{k_{p}}d^{(0)} = 
\sum_{I,J} \prod_{i \in I} d_{k_{i}}^{\mu_{i}} \prod_{j \in J} 
{\cal D}_{n}^{\mu_{j}}d^{(0)}
\label{dp-s}
\ee
where
$
I,J
$
is a partition of the set 
$
\{1,\dots,n\}
$
i.e.
$
I \cap J = \emptyset,~
I \cap J = \{1,\dots,n\}.
$
\end{thm}
As a corollary we have the generalization of (\ref{d2-F}):
\begin{thm}
We can choose the Feynman extensions such that
\be
{\partial \over \partial x_{j\mu}}
({\cal D}^{\nu_{1}}_{k_{1}}\dots {\cal D}^{\nu_{p}}_{k_{p}}d)^{F} = 
({\cal D}_{j}^{\mu}{\cal D}^{\nu_{1}}_{k_{1}}\dots {\cal D}^{\nu_{p}}_{k_{p}}d)^{F} 
- ({\cal D}_{j - 1}^{\mu}{\cal D}^{\nu_{1}}_{k_{1}}\dots {\cal D}^{\nu_{p}}_{k_{p}}d)^{F}.
\label{dp-F}
\ee
\end{thm}
{\bf Proof:} We choose convenient expressions for 
$
d^{F}, \dots, ({\cal D}_{n}^{\mu_{1}}\cdots {\cal D}^{\mu_{p}}_{n}d)^{F}
$
and define the other distributions
$
({\cal D}_{k_{1}}^{\mu_{1}}\cdots {\cal D}^{\mu_{k_{p}}}_{n}d)^{F}
$
by relations obtained by the previous ones modified with an appropriate upper
index F:
\be
({\cal D}^{\nu_{1}}_{k_{1}}\dots {\cal D}^{\nu_{p}}_{k_{p}}d)^{F} = 
\sum_{I,J} \prod_{i \in I} d_{k_{i}}^{\mu_{i}}
(\prod_{j \in J}{\cal D}_{n}^{\mu_{j}}d)^{F}.
\label{dp-sF}
\ee
The formulas from the statement are identically verified.
$\qed$

We have to consider a different type of identities verified outside
$
D_{n}
$
and see if they remain true for the Feynman extensions.
\begin{thm}
The following formulas are true:
\bea
\eta_{\rho\sigma}~
{\cal D}_{l}^{\rho}{\cal D}_{l}^{\sigma}d_{m_{1},\dots,m_{n}}^{(0)}(x_{1},\dots,x_{n})
+ m_{l}^{2}~d_{m_{1},\dots,m_{n}}^{(0)}(x_{1},\dots,x_{n})
\nonumber \\
= \delta(x_{l} - x_{l+1})~d_{m_{1},\dots,\hat{m_{l}},\dots,m_{n}}^{(0)}(x_{1},\dots,\hat{x_{l}},\dots,x_{n})
\label{shell-1}
\eea
where
$
\hat{m_{l}}
$
and
$
\hat{x_{l}}
$
means the absence of
$
m_{l}
$
and
$
x_{l}
$
respectively. If
$
n > 3
$
this relation remains true for the Feynman extensions:
\bea
\eta_{\rho\sigma}~
({\cal D}_{l}^{\rho}{\cal D}_{l}^{\sigma}d_{m_{1},\dots,m_{n}})^{F}(x_{1},\dots,x_{n})
+ m_{l}^{2}~d_{m_{1},\dots,m_{n}}^{F}(x_{1},\dots,x_{n})
\nonumber \\
= \delta(x_{l} - x_{l+1})~d_{m_{1},\dots,\hat{m_{l}},\dots,m_{n}}^{F}(x_{1},\dots,\hat{x_{l}},\dots,x_{n})
\label{shell-1F}
\eea
\label{sh-1}
\end{thm}
{\bf Proof:} The identity (\ref{shell-1}) follows immediately from the definitions. If we consider the 
Feynman extensions then we can obtain anomalies with support in
$
D_{n}
$
namely:
\bea
\eta_{\rho\sigma}~
({\cal D}_{l}^{\rho}{\cal D}_{l}^{\sigma}d_{m_{1},\dots,m_{n}})^{F}(x_{1},\dots,x_{n})
+ m_{l}^{2}~d_{m_{1},\dots,m_{n}}^{F}(x_{1},\dots,x_{n})
\nonumber \\
= \delta(x_{l} - x_{l+1})~d_{m_{1},\dots,\hat{m_{l}},\dots,m_{n}}^{F}(x_{1},\dots,\hat{x_{l}},\dots,x_{n})
+ p(\partial_{1},\dots,\partial_{n-1})\delta(x_{1},\dots,x_{n})
\eea
where 
\be
deg(p) \leq \omega(d^{(0)}) + 2 = 6 - 2n \leq - 2
\ee
for
$
n > 3,
$
so we must have 
$
p = 0
$
and (\ref{shell-1F}) is true.
$\qed$

In a similar way we prove:
\begin{thm}
The following formulas are true:
\bea
\eta_{\rho\sigma}~
{\cal D}_{k}^{\mu}{\cal D}_{l}^{\rho}{\cal D}_{l}^{\sigma}d_{m_{1},\dots,m_{n}}^{(0)}(x_{1},\dots,x_{n})
+ m_{l}^{2}~{\cal D}_{k}^{\mu}d_{m_{1},\dots,m_{n}}^{(0)}(x_{1},\dots,x_{n})
\nonumber \\
= \delta(x_{l} - x_{l+1})~{\cal D}_{k}^{\mu}d_{m_{1},\dots,\hat{m_{l}},\dots,m_{n}}^{(0)}(x_{1},\dots,\hat{x_{l}},\dots,x_{n}),
~\forall k \not= l 
\label{shell-2a}
\eea
\bea
\eta_{\rho\sigma}~
{\cal D}_{l}^{\mu}{\cal D}_{l}^{\rho}{\cal D}_{l}^{\sigma}d_{m_{1},\dots,m_{n}}^{(0)}(x_{1},\dots,x_{n})
+ m_{l}^{2}~{\cal D}_{l}^{\mu}d_{m_{1},\dots,m_{n}}^{(0)}(x_{1},\dots,x_{n})
\nonumber \\
= \partial^{\mu}\delta(x_{l} - x_{l+1})~d_{m_{1},\dots,\hat{m_{l}},\dots,m_{n}}^{(0)}(x_{1},\dots,\hat{x_{l}},\dots,x_{n})
\nonumber \\
+ \delta(x_{l} - x_{l+1})~{\cal D}_{l-1}^{\mu}d_{m_{1},\dots,\hat{m_{l}},\dots,m_{n}}^{(0)}(x_{1},\dots,\hat{x_{l}},\dots,x_{n}).
\label{shell-2b}
\eea
If
$
n > 3
$
these relations remain true for the Feynman extensions:
\bea
\eta_{\rho\sigma}~
({\cal D}_{k}^{\mu}{\cal D}_{l}^{\rho}{\cal D}_{l}^{\sigma}d_{m_{1},\dots,m_{n}})^{F}(x_{1},\dots,x_{n})
+ m_{l}^{2}~({\cal D}_{k}^{\mu}d_{m_{1},\dots,m_{n}})^{F}(x_{1},\dots,x_{n})
\nonumber \\
= \delta(x_{l} - x_{l+1})~({\cal D}_{k}^{\mu}d_{m_{1},\dots,\hat{m_{l}},\dots,m_{n}})^{F}(x_{1},\dots,\hat{x_{l}},\dots,x_{n}),
~\forall k \not= l
\label{shell-2Fa}
\eea
\bea
\eta_{\rho\sigma}~
({\cal D}_{l}^{\mu}{\cal D}_{l}^{\rho}{\cal D}_{l}^{\sigma}d_{m_{1},\dots,m_{n}})^{F}(x_{1},\dots,x_{n})
+ m_{l}^{2}~({\cal D}_{l}^{\mu}d_{m_{1},\dots,m_{n}})^{F}(x_{1},\dots,x_{n})
\nonumber \\
= \partial^{\mu}\delta(x_{l} - x_{l+1})~d_{m_{1},\dots,\hat{m_{l}},\dots,m_{n}}^{F}(x_{1},\dots,\hat{x_{l}},\dots,x_{n})
\nonumber \\
+ \delta(x_{l} - x_{l+1})~({\cal D}_{l-1}^{\mu}d_{m_{1},\dots,\hat{m_{l}},\dots,m_{n}})^{F}(x_{1},\dots,\hat{x_{l}},\dots,x_{n}).
\label{shell-2Fb}
\eea
\label{sh-2}
\end{thm}
Again we derive the absence of the anomalies from order of singularity considerations. 
The next step is more complicated.
\begin{thm}
The following formulas are true:
\bea
\eta_{\rho\sigma}~
{\cal D}_{j}^{\mu}{\cal D}_{k}^{\nu}{\cal D}_{l}^{\rho}{\cal D}_{l}^{\sigma}d_{m_{1},\dots,m_{n}}^{(0)}(x_{1},\dots,x_{n})
+ m_{l}^{2}~{\cal D}_{j}^{\mu}{\cal D}_{k}^{\nu}d_{m_{1},\dots,m_{n}}^{(0)}(x_{1},\dots,x_{n})
\nonumber \\
= \delta(x_{l} - x_{l+1})~{\cal D}_{j}^{\mu}{\cal D}_{k}^{\nu}d_{m_{1},\dots,\hat{m_{l}},\dots,m_{n}}^{(0)}(x_{1},\dots,\hat{x_{l}},\dots,x_{n}),
~\forall j,k \not= l
\label{shell-3a}
\eea
\bea
\eta_{\rho\sigma}~
{\cal D}_{j}^{\mu}{\cal D}_{l}^{\nu}{\cal D}_{l}^{\rho}{\cal D}_{l}^{\sigma}d_{m_{1},\dots,m_{n}}^{(0)}(x_{1},\dots,x_{n})
+ m_{l}^{2}~{\cal D}_{j}^{\mu}{\cal D}_{l}^{\nu}d_{m_{1},\dots,m_{n}}^{(0)}(x_{1},\dots,x_{n})
\nonumber \\
= \partial^{\nu}\delta(x_{l} - x_{l+1})~{\cal D}_{j}^{\mu}d_{m_{1},\dots,\hat{m_{l}},\dots,m_{n}}^{(0)}(x_{1},\dots,\hat{x_{l}},\dots,x_{n})
\nonumber \\
+ \delta(x_{l} - x_{l+1})~{\cal D}_{j}^{\mu}{\cal D}_{l-1}^{\nu}d_{m_{1},\dots,\hat{m_{l}},\dots,m_{n}}^{(0)}(x_{1},\dots,\hat{x_{l}},\dots,x_{n}),
~\forall j \not= l
\label{shell-3b}
\eea
\bea
\eta_{\rho\sigma}~
{\cal D}_{l}^{\mu}{\cal D}_{l}^{\nu}{\cal D}_{l}^{\rho}{\cal D}_{l}^{\sigma}d_{m_{1},\dots,m_{n}}^{(0)}(x_{1},\dots,x_{n})
+ m_{l}^{2}~{\cal D}_{l}^{\mu}{\cal D}_{l}^{\nu}d_{m_{1},\dots,m_{n}}^{(0)}(x_{1},\dots,x_{n})
\nonumber \\
= \partial^{\mu}\partial^{\nu}\delta(x_{l} - x_{l+1})~d_{m_{1},\dots,\hat{m_{l}},\dots,m_{n}}^{(0)}(x_{1},\dots,\hat{x_{l}},\dots,x_{n})
\nonumber \\
+ \partial^{\mu}\delta(x_{l} - x_{l+1})~{\cal D}_{j}^{\nu}d_{m_{1},\dots,\hat{m_{l}},\dots,m_{n}}^{(0)}(x_{1},\dots,\hat{x_{l}},\dots,x_{n})
+ (\mu \leftrightarrow \nu)
\nonumber \\
+ \delta(x_{l} - x_{l+1})~{\cal D}_{l-1}^{\mu}{\cal D}_{l-1}^{\nu}d_{m_{1},\dots,\hat{m_{l}},\dots,m_{n}}^{(0)}(x_{1},\dots,\hat{x_{l}},\dots,x_{n})
\label{shell-3c}
\eea
If
$
n > 3
$
these relations remain true for the Feynman extensions:
\bea
\eta_{\rho\sigma}~
({\cal D}_{j}^{\mu}{\cal D}_{k}^{\nu}{\cal D}_{l}^{\rho}{\cal D}_{l}^{\sigma}d_{m_{1},\dots,m_{n}})^{F}(x_{1},\dots,x_{n})
+ m_{l}^{2}~({\cal D}_{j}^{\mu}{\cal D}_{k}^{\nu}d_{m_{1},\dots,m_{n}})^{F}(x_{1},\dots,x_{n})
\nonumber \\
= \delta(x_{l} - x_{l+1})~({\cal D}_{j}^{\mu}{\cal D}_{k}^{\nu}d_{m_{1},\dots,\hat{m_{l}},\dots,m_{n}})^{F}(x_{1},\dots,\hat{x_{l}},\dots,x_{n}),
~\forall j,k \not= l
\label{shell-3Fa}
\eea
\bea
\eta_{\rho\sigma}~
({\cal D}_{j}^{\mu}{\cal D}_{l}^{\nu}{\cal D}_{l}^{\rho}{\cal D}_{l}^{\sigma}d_{m_{1},\dots,m_{n}})^{F}(x_{1},\dots,x_{n})
+ m_{l}^{2}~({\cal D}_{j}^{\mu}{\cal D}_{l}^{\nu}d_{m_{1},\dots,m_{n}})^{F}(x_{1},\dots,x_{n})
\nonumber \\
= \partial^{\nu}\delta(x_{l} - x_{l+1})~({\cal D}_{j}^{\mu}d_{m_{1},\dots,\hat{m_{l}},\dots,m_{n}})^{F}(x_{1},\dots,\hat{x_{l}},\dots,x_{n})
\nonumber \\
+ \delta(x_{l} - x_{l+1})~({\cal D}_{j}^{\mu}{\cal D}_{l-1}^{\nu}d_{m_{1},\dots,\hat{m_{l}},\dots,m_{n}})^{F}(x_{1},\dots,\hat{x_{l}},\dots,x_{n}),
~\forall j \not= l
\label{shell-3Fb}
\eea
\bea
\eta_{\rho\sigma}~
({\cal D}_{l}^{\mu}{\cal D}_{l}^{\nu}{\cal D}_{l}^{\rho}{\cal D}_{l}^{\sigma}d_{m_{1},\dots,m_{n}})^{F}(x_{1},\dots,x_{n})
+ m_{l}^{2}~({\cal D}_{l}^{\mu}{\cal D}_{l}^{\nu}d_{m_{1},\dots,m_{n}})^{F}(x_{1},\dots,x_{n})
\nonumber \\
= \partial^{\mu}\partial^{\nu}\delta(x_{l} - x_{l+1})~d_{m_{1},\dots,\hat{m_{l}},\dots,m_{n}}^{F}(x_{1},\dots,\hat{x_{l}},\dots,x_{n})
\nonumber \\
+ \partial^{\mu}\delta(x_{l} - x_{l+1})~({\cal D}_{j}^{\nu}d_{m_{1},\dots,\hat{m_{l}},\dots,m_{n}})^{F}(x_{1},\dots,\hat{x_{l}},\dots,x_{n})
+ (\mu \leftrightarrow \nu)
\nonumber \\
+ \delta(x_{l} - x_{l+1})~({\cal D}_{l-1}^{\mu}{\cal D}_{l-1}^{\nu}d_{m_{1},\dots,\hat{m_{l}},\dots,m_{n}})^{F}(x_{1},\dots,\hat{x_{l}},\dots,x_{n}).
\label{shell-3Fc}
\eea
\label{sh-3}
\end{thm}
{\bf Proof:} The relations (\ref{shell-3a}) - (\ref{shell-3c}) are derived by direct computations. When we go to the 
Feynman extensions we get anomalies
\bea
\eta_{\rho\sigma}~
({\cal D}_{j}^{\mu}{\cal D}_{k}^{\nu}{\cal D}_{l}^{\rho}{\cal D}_{l}^{\sigma}d_{m_{1},\dots,m_{n}})^{F}(x_{1},\dots,x_{n})
+ m_{l}^{2}~({\cal D}_{j}^{\mu}{\cal D}_{k}^{\nu}d_{m_{1},\dots,m_{n}})^{F}(x_{1},\dots,x_{n})
\nonumber \\
= \delta(x_{l} - x_{l+1})~({\cal D}_{j}^{\mu}{\cal D}_{k}^{\nu}d_{m_{1},\dots,\hat{m_{l}},\dots,m_{n}})^{F}(x_{1},\dots,\hat{x_{l}},\dots,x_{n})
+ a^{\mu\nu}_{jk}(x_{1},\dots,x_{n}),
~\forall j,k \not= l
\eea
\bea
\eta_{\rho\sigma}~
({\cal D}_{j}^{\mu}{\cal D}_{l}^{\nu}{\cal D}_{l}^{\rho}{\cal D}_{l}^{\sigma}d_{m_{1},\dots,m_{n}})^{F}(x_{1},\dots,x_{n})
+ m_{l}^{2}~({\cal D}_{j}^{\mu}{\cal D}_{l}^{\nu}d_{m_{1},\dots,m_{n}})^{F}(x_{1},\dots,x_{n})
\nonumber \\
= \partial^{\nu}\delta(x_{l} - x_{l+1})~({\cal D}_{j}^{\mu}d_{m_{1},\dots,\hat{m_{l}},\dots,m_{n}})^{F}(x_{1},\dots,\hat{x_{l}},\dots,x_{n})
\nonumber \\
+ \delta(x_{l} - x_{l+1})~({\cal D}_{j}^{\mu}{\cal D}_{l-1}^{\nu}d_{m_{1},\dots,\hat{m_{l}},\dots,m_{n}})^{F}(x_{1},\dots,\hat{x_{l}},\dots,x_{n})
+ a^{\mu\nu}_{jl}(x_{1},\dots,x_{n}),
~\forall j \not= l
\eea
\bea
\eta_{\rho\sigma}~
({\cal D}_{l}^{\mu}{\cal D}_{l}^{\nu}{\cal D}_{l}^{\rho}{\cal D}_{l}^{\sigma}d_{m_{1},\dots,m_{n}})^{F}(x_{1},\dots,x_{n})
+ m_{l}^{2}~({\cal D}_{l}^{\mu}{\cal D}_{l}^{\nu}d_{m_{1},\dots,m_{n}})^{F}(x_{1},\dots,x_{n})
\nonumber \\
= \partial^{\mu}\partial^{\nu}\delta(x_{l} - x_{l+1})~d_{m_{1},\dots,\hat{m_{l}},\dots,m_{n}}^{F}(x_{1},\dots,\hat{x_{l}},\dots,x_{n})
\nonumber \\
+ \partial^{\mu}\delta(x_{l} - x_{l+1})~({\cal D}_{j}^{\nu}d_{m_{1},\dots,\hat{m_{l}},\dots,m_{n}})^{F}(x_{1},\dots,\hat{x_{l}},\dots,x_{n})
+ (\mu \leftrightarrow \nu)
\nonumber \\
+ \delta(x_{l} - x_{l+1})~({\cal D}_{l-1}^{\mu}{\cal D}_{l-1}^{\nu}d_{m_{1},\dots,\hat{m_{l}},\dots,m_{n}})^{F}(x_{1},\dots,\hat{x_{l}},\dots,x_{n})
+ a^{\mu\nu}_{ll}(x_{1},\dots,x_{n}).
\eea
Here the anomalies have the structure
\be
a^{\mu\nu}_{jk}(x_{1},\dots,x_{n}) = p^{\mu\nu}_{jk}(\partial_{1},\dots,\partial_{n-1})\delta(x_{1},\dots,x_{n})
\ee
with
\be
deg(p^{\mu\nu}_{jk}) \leq 4 + \omega(d^{(0)}) = 8 - 4n
\ee
and can be non-trivial. They must also satisfy the symmetry property
\be
a^{\mu\nu}_{jk} = a^{\nu\mu}_{kj}.
\ee

Now we apply the operator
$
{\partial \over \partial x_{l\mu}}
$
on the relations from the preceding theorem and use the identities (\ref{dp-F}). After some computations we derive
that the anomalies
$
a^{\mu\nu}_{jk}
$
verify 
\be
a^{\mu\nu}_{jk} = a^{\mu\nu}_{j-1,k}
\ee
so do not depend on 
$
k,j
$
i.e.
\be
a^{\mu\nu}_{jk} = a^{\mu\nu}.
\ee
We have the generic expression
\be
a^{\mu\nu}(x_{1},\dots,x_{n}) = p^{\mu\nu}(\partial_{1},\dots,\partial_{n-1})\delta(x_{1},\dots,x_{n})
\ee
with
\be
deg(p^{\mu\nu}) \leq 4 + \omega(d^{(0)}) = 8 - 4n.
\ee
One the other hand we can make the redefinitions
\bea
({\cal D}_{j}^{\mu}{\cal D}_{k}^{\nu}{\cal D}_{l}^{\rho}{\cal D}_{m}^{\sigma}d_{m_{1},\dots,m_{n}})^{F}
\rightarrow
\nonumber \\
({\cal D}_{j}^{\mu}{\cal D}_{k}^{\nu}{\cal D}_{l}^{\rho}{\cal D}_{m}^{\sigma}d_{m_{1},\dots,m_{n}})^{F}
+ p^{\mu\nu\rho\sigma}(\partial_{1},\dots,\partial_{n-1})\delta(x_{1},\dots,x_{n})
\eea
with
\be
deg(p^{\mu\nu\rho\sigma}) \leq  8 - 4n
\ee
without affecting the relations (\ref{dp-F}). This redefinitions change the anomaly:
\be
a^{\mu\nu} \rightarrow a^{\mu\nu} - \eta_{\rho\sigma}~p^{\mu\nu\rho\sigma}
\ee
so if we choose conveniently
$
p^{\mu\nu\rho\sigma}
$
we can make null the anomalies
$
a^{\mu\nu} 
$
and this proves the theorem.
$\qed$

We can continue by induction and consider more derivatives
$
{\cal D}_{j_{1}}^{\mu_{1}}\dots,{\cal D}_{j_{p}}^{\mu_{p}},~p > 2
$
and obtain the same conclusion: the identities obtained outside
$
D_{n}
$
are preserved by the Feynman extensions.

Now we have a result similar to theorem \ref{sh-1}:
\begin{thm}
The following formulas are true for
$
k < l:
$
\bea
\eta_{\mu\nu}~\eta_{\rho\sigma}~
{\cal D}_{k}^{\mu}{\cal D}_{k}^{\nu}{\cal D}_{l}^{\rho}{\cal D}_{l}^{\sigma}d_{m_{1},\dots,m_{n}}^{(0)}(x_{1},\dots,x_{n})
- m_{k}^{2}~ m_{l}^{2}~d_{m_{1},\dots,m_{n}}^{(0)}(x_{1},\dots,x_{n})
\nonumber \\
= - [ m_{k}^{2}~\delta(x_{l} - x_{l+1})~d_{m_{1},\dots,\hat{m_{l}},\dots,m_{n}}^{(0)}(x_{1},\dots,\hat{x_{l}},\dots,x_{n})
+ ( k \leftrightarrow l) ]
\nonumber \\
+ \delta(x_{k} - x_{k+1})~\delta(x_{l} - x_{l+1})~
d_{m_{1},\dots,\hat{m_{k}},\dots,\hat{m_{l}},\dots,m_{n}}^{(0)}(x_{1},\dots,\hat{x_{k}},\dots,\hat{x_{l}},\dots,x_{n}).
\label{shell-4}
\eea
If
$
n > 4
$
this relation remains true for the Feynman extensions:
\bea
\eta_{\mu\nu}~\eta_{\rho\sigma}~
({\cal D}_{k}^{\mu}{\cal D}_{k}^{\nu}{\cal D}_{l}^{\rho}{\cal D}_{l}^{\sigma}d_{m_{1},\dots,m_{n}})^{F}(x_{1},\dots,x_{n})
- m_{k}^{2}~ m_{l}^{2}~d_{m_{1},\dots,m_{n}}^{F}(x_{1},\dots,x_{n})
\nonumber \\
= - [ m_{k}^{2}~\delta(x_{l} - x_{l+1})~d_{m_{1},\dots,\hat{m_{l}},\dots,m_{n}}^{F}(x_{1},\dots,\hat{x_{l}},\dots,x_{n})
+ ( k \leftrightarrow l) ]
\nonumber \\
+ \delta(x_{k} - x_{k+1})~\delta(x_{l} - x_{l+1})~
d_{m_{1},\dots,\hat{m_{k}},\dots,\hat{m_{l}},\dots,m_{n}}^{F}(x_{1},\dots,\hat{x_{k}},\dots,\hat{x_{l}},\dots,x_{n}).
\label{shell-4F}
\eea
\label{sh2}
\end{thm}
{\bf Proof:} The first formula follows from direct computations. When we extend this formula to
$
D_{n}
$
we can have anomalies
\be
a_{kl}(x_{1},\dots,x_{n}) = p_{kl}(\partial_{1},\dots,\partial_{n-1})\delta(x_{1},\dots,x_{n})
\ee
with
\be
deg(p_{kl}) \leq 4 + \omega(d^{(0)}) = 8 - 2n \leq - 2
\ee
for
$
n > 4
$
so in fact there are no anomalies.
$\qed$

Now we can extend the formula recursively as before adding supplementary derivatives
$
{\cal D}_{j_{1}}^{\mu_{1}}\dots,{\cal D}_{j_{p}}^{\mu_{p}},~p > 2
$
and obtain the same conclusion:  the identities obtained outside
$
D_{n}
$
are preserved by the Feynman extensions.

Finally, we can analyze in the same way the general case when there are $p$ contractions
$
\eta_{\mu_{1}\nu_{1}}\dots,\eta_{\mu_{p}\nu_{p}}
$
and $q$ ``free" derivatives
$
{\cal D}_{j_{1}}^{\mu_{1}}\dots,{\cal D}_{j_{q}}^{\mu_{q}};
$
because 
$
2p \leq n 
$
we arrive at the same conclusion as above.

We consider from now on the Yang-Mills case. 
One can prove \cite{cohomology} that the tree contributions can produce anomalies only for 
$
n \leq 3.
$
Suppose that we have eliminated the anomalies of one-loop graphs in order
$
n = 3
$
of the perturbation theory. Then we can use induction to extend the result for 
one-loop contributions in an arbitrary order of the perturbation theory.

\begin{thm}
The one-loop contributions do not produce anomalies in orders
$
n > 4
$
of the perturbation theory.
\label{gauge3}
\end{thm}
{\bf Proof:} We proceed by induction. We denote by
$
T^{I_{1},\dots,I_{p}}_{(l)}, l =1,2,,,
$
the contribution associated with $l$ loop graphs from the chronological products.
Suppose that the assertion is true for 
$
1,\dots,n-1
$
i.e. we have
\be
sT^{I_{1},\dots,I_{p}}_{(1)} = 0,~p = 1,\dots,n - 1.
\ee
One the other hand the identity holds for three contributions also:
\be
sT^{I_{1},\dots,I_{p}}_{(0)} = 0,~\forall p
\ee
as we have said above. Outside the set 
$
D_{n}
$
we can use the preceding formula to prove
\be
sT^{I_{1},\dots,I_{n}}_{(1)} = 0,\quad \forall (x_{1},\dots,x_{n}) \not\in D_{n}
\label{generic-n}
\ee
so it remains to see if we can extend this identity to the whole space. 

These loop contribution to 
$
T^{I_{1},\dots,I_{n}}
$
are sums of contributions of the type
\bea
{\cal D}^{\mu_{1}}_{j_{1}}\dots {\cal D}^{\mu_{p}}_{j_{p}}
d^{(0)}_{m_{1},\dots,m_{n}}(x_{1},\dots,x_{n})~W(x_{1},\dots,x_{n}),~p \leq n
\label{gauge-n}
\eea
where $W$ are Wick monomials. These expressions come with various numerical
coefficients $t$ (which are in fact Lorentz tensors). This follows from the
limitations of the Yang-Mills model: we have at most a derivative in the
interaction Lagrangian, so we have at most $n$ derivatives 
$
{\cal D}
$
on $d$. If one expands (\ref{generic-n}) using the explicit forms,
one is reduced to identities of the type (\ref{dp}) and (\ref{shell-1}), 
(\ref{shell-2a}), (\ref{shell-2b}), (\ref{shell-3a}) - (\ref{shell-3c}), (\ref{shell-4}).
So we will have
\be
sT^{I_{1},\dots,I_{n}}_{(1)} = 0
\label{generic-nF}
\ee
if these identities can be extended without anomalies. But this is exactly what we have
proved above.
$\qed$

We point out that the origin of the anomalies is the fact
that the operation of extension of distributions and the operation of taking
the contraction with the Minkowski metric
$
\eta_{\cdot \cdot}
$
do {\bf not} commute and the difference is a potential anomaly. For 
$
n > 4
$
the preceding theorem shows that such anomalies do not appear. It remains to study
lower orders of perturbation theory.
\newpage

\section{Second Order Anomalies\label{second}}

In second order we have some typical distributions. 
We remind the fact that the Pauli-Villars distribution is defined by
\be
D_{m}(x) = D_{m}^{(+)}(x) + D_{m}^{(-)}(x)
\ee
where 
\be
D_{m}^{(\pm)}(x) = \pm {i \over (2\pi)^{3}}~
\int dp e^{i p\cdot x} \theta(\pm p_{0}) \delta(p^{2} - m^{2})
\ee
such that
\be
D^{(-)}(x) = - D^{(+)}(- x).
\ee

This distribution has causal support. In fact, it can be causally split
(uniquely) into an
advanced and a retarded part:
\be
D = D^{\rm adv} - D^{\rm ret}
\ee
and then we can define the Feynman propagator and antipropagator
\be
D^{F} = D^{\rm ret} + D^{(+)}, \qquad \bar{D}^{F} = D^{(+)} - D^{\rm adv}.
\ee
All these distributions have singularity order
$
\omega(D) = -2
$.

For  one-loop contributions in the second order we need the basic distributions
\be
d^{(2)}_{D_{1},D_{2}}(x) \equiv
{1 \over 2}~[ D_{1}^{(+)}(x)~D_{2}^{(+)}(x) - D_{1}^{(-)}(x)~D_{2}^{(-)}(x) ]
\ee
where
$
D_{j} = D_{m_{j}}
$
which also with causal support. This expression is linear in
$
D_{1}
$
and
$
D_{2}
$.
We will also use the notation
\be
d_{12} = d^{(2)}_{D_{1},D_{2}}
\ee
and when no confusion about the distributions
$
D_{j} = D_{m_{j}}
$
can appear, we skip all indexes altogether. The causal split
\be
d_{12} = d_{12}^{adv} - d_{12}^{ret}
\ee
is not unique because
$
\omega(d_{12}) = 0
$
so we make the redefinitions
\be
d_{12}^{adv(ret)}(x) \rightarrow d_{12}^{adv(ret)}(x) + c~\delta(x)
\ee
without affecting the support properties and the order of singularity.
The corresponding Feynman propagators can be defined as above and will be
denoted as
$
d_{12}^{F}
$.
Another way to construct them is to define for
$
x \not= 0
$
the distribution
\be
d^{(0)}_{12}(x) \equiv {1\over 2}~D_{1}^{F}(x)~D_{2}^{F}(x)
\ee
and to extend it to the whole domain using a standard result in distribution
theory (see the preceding Section). 

We will consider the case
$
D_{1} = D_{2} = D_{m}
$
and determine its Fourier transform; by direct computations it can be obtained
that
\be
\tilde{d}_{m,m}(k)
\equiv {1 \over (2\pi)^{2}} \int dx~ e^{i k\cdot x} d_{m,m}(x)
= - {1 \over 8 (2\pi)^{3}}~\varepsilon(k_{0})~\theta(k^{2} - m^{2}) 
\sqrt{1 - {4 m^{2} \over k^{2}}}.
\label{d-mm}
\ee

We also define the distributions
\bea
d^{\mu\nu}(x) = 
D^{(+)}_{m}(x) \partial^{\mu}\partial^{\nu}D^{(+)}_{m}(x) 
- D^{(-)}_{m}(x) \partial^{\mu}\partial^{\nu}D^{(-)}_{m}(x) 
\nonumber \\
f^{\mu\nu}(x) = 
\partial^{\mu}D^{(+)}_{m}(x) \partial^{\nu}D^{(+)}_{m}(x) 
- \partial^{\mu}D^{(-)}_{m}(x) \partial^{\nu}D^{(-)}_{m}(x) 
\eea

Performing a Fourier transform we can obtain the formula
\be
d^{\mu\nu}(x) = {2\over 3} 
\left(\partial^{\mu}\partial^{\nu} - 
{1\over 4}\eta^{\mu\nu}\square\right)d_{m,m}(x)
- {2 m^{2}\over 3}
(\partial^{\mu}\partial^{\nu} - \eta^{\mu\nu}\square)
d^{\prime}_{m,m}(x)
\ee
where we define the distribution
$
d^{\prime}_{m,m}(x)
$
through its Fourier transform:
\be
\tilde{d^{\prime}}_{m,m}(k) = {1\over k^{2}}~\tilde{d}_{m,m}(k).
\ee
This distribution also has causal support and it verifies
\be
\square d^{\prime}_{m,m} = - d_{m,m}.
\label{d-prime}
\ee
It can be proved that the central causal splitting preserves this relation. The
distribution
\be
f^{\mu\nu} = 2 {\cal D}_{1}^{\mu}{\cal D}_{2}^{\nu}d
\ee
is simply obtained as
\be
f^{\mu\nu} = \partial^{\mu}\partial^{\nu}d_{m,m} - d^{\mu\nu}.
\ee

The dominant contribution can produce anomalies of canonical dimension $5$ and
the super-renormalizable contributions can produce anomalies of canonical
dimension at most $4$. We investigate the dominant anomaly.

We now consider the one-loop contributions 
$
D_{(1)}^{IJ}(x,y)
$
from
$
D^{IJ}(x,y)
$
and we write for every mass $m$ in the game
\be
D_{m} = D_{M} + ( D_{M} - D_{m})
\label{split}
\ee
In this way we split 
$
D_{(1)}^{IJ}(x,y)
$
into a dominant contribution 
$
D_{\rm dominant}^{IJ}(x,y)
$
where everywhere
$
D_{m} \mapsto D_{M}
$
and a contribution where at least one factor
$
D_{m}
$
is replaced by the difference
$
D_{m} - D_{M}
$.
Because we have
\be
\omega(D_{m} - D_{M}) = -4
\ee
the second contribution will be super-renormalizable. The dominant contribution can produce anomalies of
maximal dimension 
$
\omega({\cal A}) = 5
$
and rest will produce anomalies with canonical dimension 
$
\omega({\cal A}) \leq 4.
$

We now consider the dominant contribution. By direct computations we obtain
\be
D_{\rm dominant}^{[\mu\nu]\emptyset}(x,y) = 0
\ee
\be
D_{\rm dominant}^{[\mu][\nu]}(x,y) = (\partial^{\mu}\partial^{\nu} - \eta^{\mu\nu}
\square)d_{M,M}(x - y) 
\tilde{g}_{ab} u_{a}(x) u_{b}(y)
\ee
\bea
D_{\rm dominant}^{[\mu] \emptyset}(x,y) = (\partial^{\mu}\partial^{\nu} -
\eta^{\mu\nu}
\square) d_{M,M}(x - y) 
\tilde{g}_{ab} u_{a}(x) v_{b\nu}(y)
\nonumber \\
+ \partial_{\nu} d_{M,M}(x - y) g_{ab} [ F^{\mu\nu}_{a}(x) u_{b}(y) - u_{a}(x)
F^{\mu\nu}_{b}(y) ]
\eea
\be
D_{\rm dominant}^{\emptyset [\mu]}(x,y) = - D_{(1)0}^{[\mu]\emptyset}(y,x)
\ee
\bea
D_{\rm dominant}^{\emptyset \emptyset}(x,y) = (\partial^{\mu}\partial^{\nu} -
\eta^{\mu\nu}
\square)d_{M,M}(x - y) 
\tilde{g}_{ab} v_{a\mu}(x) v_{b\nu}(y)
\nonumber \\
+ \partial_{\mu} d_{M,M}(x - y)
g_{ab} [ - F^{\mu\nu}_{a}(x) v_{b\nu}(y) +  \partial^{\mu}\tilde{u}_{a}(x)
u_{b}(y)
 + v_{a\nu}(x) F^{\mu\nu}_{b}(y) - u_{a}(x)  \partial^{\mu}\tilde{u}_{b}(y) ]
\nonumber \\
- d_{M,M}(x - y)g_{ab} F^{\mu\nu}_{a}(x) F_{b\mu\nu}(y) 
\nonumber \\
+ \partial_{\mu} d_{M,M}(x - y)
g^{(3)}_{ab} [ \Phi_{a}(x) \partial^{\mu}\Phi_{b}(y) -
\partial^{\mu}\Phi_{a}(x) \Phi_{b}(y) ]
- 2 d_{M,M}(x - y) g^{(3)}_{ab} \partial^{\mu}\Phi_{a}(x)
\partial_{\mu}\Phi_{b}(y)
\nonumber \\
- i \partial_{\mu} d_{M,M}(x - y)
[ \bar{\Psi}(x) A_{\epsilon} \otimes \gamma^{\mu}\gamma_{\epsilon}\Psi(y) 
- \bar{\Psi}(y) A_{\epsilon} \otimes \gamma^{\mu}\gamma_{\epsilon}\Psi(x)]
\nonumber \\
 + \square  d_{M,M}(x - y) g^{(4)}_{ab} \Phi_{a}(x) \Phi_{b}(y)
\eea
where we have defined some bilinear combinations in the constants appearing in
the 
interaction Lagrangian:
\bea
g_{ab} = f_{pqa} f_{pqb}
\qquad
g^{(1)}_{ab} = f^{\prime}_{pqa} f^{\prime}_{pqb} 
\qquad
g^{(2)}_{ab} = \sum_{\epsilon} Tr(t^{\epsilon}_{a} t^{\epsilon}_{b})
\qquad
g^{(3)}_{ab} = f^{\prime}_{apq} f^{\prime}_{bpq} 
\nonumber \\
g^{(4)}_{ab} = 2 \sum_{\epsilon} Tr(s^{\epsilon}_{a} s^{- \epsilon}_{b})
\qquad
\tilde{g}_{ab} \equiv {1\over 3}~(2 g_{ab} + g^{(1)}_{ab} + 4 g^{(2)}_{ab})
\qquad
A_{\epsilon} = \sum_{a} ( 2 t^{\epsilon}_{a} t^{\epsilon}_{a} +
s^{-\epsilon}_{a} s^{\epsilon}_{a}).
\eea

It is easy to see that the substitution
\be
d_{M,M}(x - y) \rightarrow d_{M,M}^{F}(x - y) 
\ee
gives the dominant contribution to the chronological product and does not produce anomalies.
So only anomalies of lower dimension can appear.

\newpage
\section{Third Order Causal Distributions\label{third}}

For the triangle one-loop contributions in the third order we give an alternative construction
of the relevant Feynman distributions. First, we take
$
D_{j} = D_{m_{j}}, j = 1,2,3
$
and define
\bea
d_{D_{1},D_{2},D_{3}}^{(3)}(x,y,z) \equiv \bar{D}^{F}_{3}(x - y) 
[ D^{(-)}_{2}(z - x) D^{(+)}_{1}(y - z) - D^{(+)}_{2}(z - x) D^{(-)}_{1}(y - z)
]
\nonumber \\
+ D^{F}_{1}(y - z) 
[ D^{(-)}_{3}(x - y) D^{(+)}_{2}(z - x) - D^{(+)}_{3}(x - y) D^{(-)}_{2}(z - x)
]
\nonumber \\
+ D^{F}_{2}(z - x) 
[ D^{(-)}_{1}(y - z) D^{(+)}_{3}(x - y) - D^{(+)}_{1}(y - z) D^{(-)}_{3}(x - y)
]
\eea
which also with causal support; indeed we have the alternative forms
\bea
d_{D_{1},D_{2},D_{3}}^{(3)}(x,y,z) = - D^{\rm ret}_{3}(x - y) 
[ D^{(-)}_{2}(z - x) D^{(+)}_{1}(y - z) - D^{(+)}_{2}(z - x) D^{(-)}_{1}(y - z)
]
\nonumber \\
+ D^{\rm adv}_{1}(y - z) 
[ D^{(-)}_{3}(x - y) D^{(+)}_{2}(z - x) - D^{(+)}_{3}(x - y) D^{(-)}_{2}(z - x)
]
\nonumber \\
+ D^{\rm adv}_{2}(z - x) 
[ D^{(-)}_{1}(y - z) D^{(+)}_{3}(x - y) - D^{(+)}_{1}(y - z) D^{(-)}_{3}(x - y)
]
\eea
and
\bea
d_{D_{1},D_{2},D_{3}}^{(3)}(x,y,z) = - D^{\rm adv}_{3}(x - y) 
[ D^{(-)}_{2}(z - x) D^{(+)}_{1}(y - z) - D^{(+)}_{2}(z - x) D^{(-)}_{1}(y - z)
]
\nonumber \\
+ D^{\rm ret}_{1}(y - z) 
[ D^{(-)}_{3}(x - y) D^{(+)}_{2}(z - x) - D^{(+)}_{3}(x - y) D^{(-)}_{2}(z - x)
]
\nonumber \\
+ D^{\rm ret}_{2}(z - x) 
[ D^{(-)}_{1}(y - z) D^{(+)}_{3}(x - y) - D^{(+)}_{1}(y - z) D^{(-)}_{3}(x - y)
]
\eea
from which it follows that 
$
d_{D_{1},D_{2},D_{3}}^{(3)}(x,y,z)
$
is null outside the causal cone
$
\{ (x,y,z) | x - z \in V^{+}, y - z \in V^{+}  \} \cup 
\{ (x,y,z) | x - z \in V^{-}, y - z \in V^{-}  \}
$. 
These distributions have the singularity order
$
\omega(d_{D_{1},D_{2},D_{3}}^{(3)}) = - 2
$.

As in the previous Section we use the alternative notation
\be
d_{123} \equiv d_{D_{1},D_{2},D_{3}}^{(3)}
\ee
and when there is no ambiguity about the distributions
$
D_{j}
$
we simply denote
$
d = d_{123}
$.
There are some associated distributions obtained from
$
d_{D_{1},D_{2},D_{3}}(x,y,z)
$
applying derivatives on the factors
$
D_{j} = D_{m_{j}}, j = 1,2,3
$.
For instance we denote
\bea
{\cal D}^{1}_{\alpha}d_{D_{1},D_{2},D_{3}} \equiv
d_{\partial_{\alpha}D_{1},D_{2},D_{3}},\quad
{\cal D}^{2}_{\alpha}d_{D_{1},D_{2},D_{3}} \equiv
d_{D_{1},\partial_{\alpha}D_{2},D_{3}},\quad
{\cal D}^{3}_{\alpha}d_{D_{1},D_{2},D_{3}} \equiv
d_{D_{1},D_{2},\partial_{\alpha}D_{3}},
\eea
and so on for more derivatives
$
\partial_{\alpha}
$
distributed in an arbitrary way on the factors
$
D_{j} = D_{m_{j}}, j = 1,2,3
$.
We mention the fact that the operators
$
{\cal D}^{j}_{\alpha}, j = 1,2,3
$
are commutative but they are not derivation operators: they do not verify
Leibniz rule. We note that we have:
\bea
{\partial \over \partial x_{\mu}}d = 
( {\cal D}_{3}^{\mu} - {\cal D}_{2}^{\mu})d, \quad
{\partial \over \partial y_{\mu}}d =
( {\cal D}_{1}^{\mu} - {\cal D}_{3}^{\mu})d, \quad
{\partial \over \partial z_{\mu}}d =
( {\cal D}_{2}^{\mu} - {\cal D}_{1}^{\mu})d.
\label{d123-p}
\eea

It is known that these distributions can be causally split in such a way that
the order of singularity, translation invariance and Lorentz covariance are
preserved. The same will be true for the corresponding Feynman distributions.
Because 
$
\omega(d_{123}) = - 2
$
and
$
\omega({\cal D}_{i}^{\mu}d_{123}) = - 1
$
the corresponding advanced, retarded and Feynman distributions are unique. For
more derivatives we have some freedom of redefinition. There is an alternative
way to define these distributions presented in Section \ref{one-loop}. 

As in the previous Section, let us consider the case
$
D_{1} = D_{2} = D_{3} = D_{m},~m > 0
$
and study the corresponding distribution
$
d_{m,m,m}.
$
We consider it as distribution in two variables
$
X \equiv x - z,\quad Y \equiv y - z
$
and we will need its Fourier transform. The computation is essentially done in
\cite{Sc1} and gives the following formula:
\be
\tilde{d}_{m,m,m}(p,q) = {1\over 8 (2\pi)^{5}} {1 \over \sqrt{N}}~
[\epsilon(p_{0}) \theta(p^{2} - 4 m^{2})~ln_{1}
+ \epsilon(q_{0}) \theta(q^{2} - 4 m^{2})~ln_{2} 
+ \epsilon(P_{0}) \theta(P^{2} - 4 m^{2})~ln_{3} ]
\label{d-mmm}
\ee
where
\bea
ln_{1} \equiv ln\left({P\cdot q + \sqrt{N (1 - 4 m^{2}/p^{2})} \over
P\cdot q - \sqrt{N (1 - 4 m^{2}/p^{2})}}\right)
\nonumber \\
ln_{2} \equiv ln\left({P\cdot p + \sqrt{N (1 - 4 m^{2}/q^{2})} \over
P\cdot p - \sqrt{N (1 - 4 m^{2}/q^{2})}}\right)
\nonumber \\
ln_{3} \equiv ln\left({- p\cdot q + \sqrt{N (1 - 4 m^{2}/P^{2})} \over
- p\cdot q - \sqrt{N (1 - 4 m^{2}/P^{2})}}\right)
\eea
with the notations
$
P = p + q
$
and
$
N \equiv (p\cdot q)^{2} - p^{2} q^{2}.
$

Now we define the distributions with causal support
\bea
f_{1}(x,y,z) = \delta(y - z)~d_{m,m}(x - y)
\nonumber \\
f_{2}(x,y,z) = \delta(z - x)~d_{m,m}(y - z)
\nonumber \\
f_{3}(x,y,z) = \delta(x - y)~d_{m,m}(y - z)
\eea
which do appear when considering 1-particle reducible graphs. We consider
them (as before) as distributions in two variables
$
X \equiv x - z,~Y \equiv y - z
$
and the Fourier transforms are:
\be
\tilde{f}_{1}(p,q) = {1 \over (2\pi)^{2}}~\tilde{d}_{m,m}(p),\quad
\tilde{f}_{2}(p,q) = {1 \over (2\pi)^{2}}~\tilde{d}_{m,m}(q),\quad
\tilde{f}_{3}(p,q) = {1 \over (2\pi)^{2}}~\tilde{d}_{m,m}(P)
\label{f}
\ee

Similarly we define
\bea
f_{1}^{\prime}(x,y,z) = \delta(y - z)~d_{m,m}^{\prime}(x - y)
\nonumber \\
f_{2}^{\prime}(x,y,z) = \delta(z - x)~d_{m,m}^{\prime}(y - z)
\nonumber \\
f_{3}^{\prime}(x,y,z) = \delta(x - y)~d_{m,m}^{\prime}(y - z).
\eea

Let us denote for simplicity 
\be
{\cal K}_{j} = \eta_{\rho\sigma}~{\cal D}_{j}^{\rho}{\cal D}_{j},~
j = 1,2,3
\ee
the derivative operators
\be
\partial^{\mu}_{1} \equiv {\partial \over \partial X_{\mu}},\quad
\partial^{\mu}_{2} \equiv {\partial \over \partial Y_{\mu}},\quad
\partial^{\mu}_{3} \equiv - \partial^{\mu}_{1} - \partial^{\mu}_{2},\quad
\square_{j} \equiv \partial_{j}\cdot\partial_{j}
\ee
and we have by direct computation:
\begin{thm}
The following relations are true
\be
({\cal K}_{l} + m^{2})d_{m,m,m} = 2~f_{l},
\label{shell1}
\ee
\be
{\cal D}_{j}^{\mu}({\cal K}_{l} + m^{2})d_{m,m,m} = f^{\mu}_{jl},
\label{shell2}
\ee
\be
{\cal D}_{j}^{\mu}{\cal D}_{k}^{\nu}({\cal K}_{l} + m^{2})d_{m,m,m} =
f^{\mu\nu}_{jkl} - {2 m^{2} \over 3}~C^{\mu\nu}f^{\prime}_{l}
\label{shell3}
\ee
\be
{\cal K}_{j}{\cal K}_{l}d_{m,m,m} = m^{4}~d_{m,m,m} 
- 2 m^{2}(f_{j} + f_{l})
\label{shell4}
\ee
\be
{\cal D}_{j}\cdot{\cal D}_{k}{\cal K}_{l} = f_{jkl} - 2 m^{2} f_{l}
\label{shell5}
\ee
where
\bea
f^{\mu}_{11} = (\partial^{\mu}_{1} + 2 \partial^{\mu}_{2})f_{1},\quad
f^{\mu}_{21} = - \partial^{\mu}_{1}f_{1},\quad
f^{\mu}_{31} = \partial^{\mu}_{1}f_{1}
\nonumber \\
f^{\mu\nu}_{221} = f^{\mu\nu}_{331} = A^{\mu\nu}_{1}f_{1},\quad
f^{\mu\nu}_{231} = - B^{\mu\nu}_{1}f_{1},
\nonumber \\
f^{\mu\nu}_{131} = 
(\partial^{\nu}_{1}\partial^{\mu}_{2} + A^{\mu\nu}_{1})f_{1},\quad
f^{\mu\nu}_{121} = 
- (\partial^{\nu}_{1}\partial^{\mu}_{2} + B^{\mu\nu}_{1})f_{1},
\nonumber \\
f^{\mu\nu}_{111} = 
(\partial^{\nu}_{1}\partial^{\mu}_{2} + \partial^{\mu}_{1}\partial^{\nu}_{2}
+ 2 \partial^{\mu}_{2}\partial^{\nu}_{2} + A^{\mu\nu}_{1})f_{1}
\nonumber \\
f_{231} = - \square_{1}f_{1},\quad
f_{131} = - \partial_{1}\cdot\partial_{2}f_{1},\quad
f_{121} = - \partial_{1}\cdot\partial_{3}f_{1}
\eea
and the rest by circular permutations. Here we have defined
\bea
A^{\mu\nu}_{j} \equiv {2 \over 3} \left(\partial^{\mu}_{j}\partial^{\mu}_{j}
- {1\over 4} \eta^{\mu\nu}~\square_{j}\right)
\nonumber \\
B^{\mu\nu}_{j} \equiv {1 \over 3} \left(\partial^{\mu}_{j}\partial^{\mu}_{j}
+ {1\over 2} \eta^{\mu\nu}~\square_{j}\right)
\nonumber \\
C^{\mu\nu}_{j} \equiv (\partial^{\mu}_{j}\partial^{\mu}_{j}
- \eta^{\mu\nu}~\square_{j}).
\eea
\label{shell}
\end{thm}

The anomalies are produced by the causal splitting of these relations. To
obtain these anomalies we have to determine the Fourier transforms of the
associated distributions.

First we consider the distributions
\be
d^{\mu}_{j} \equiv {\cal D}_{j}^{\mu}d_{m,m,m}
\ee

From Lorentz covariance considerations the Fourier transform should be of the
form:
\be
\tilde{d}^{\mu}_{j}(p,q) = - i~[p^{\mu}~\tilde{A}_{j}(p,q) +
q^{\mu}~\tilde{B}_{j}(p,q)]
\label{d11}
\ee
where the scalar functions
$
\tilde{A}_{j}
$
and
$
\tilde{B}_{j}
$
depend in fact only on the Lorentz invariants:
$
p^{2}, q^{2}, p\cdot q.
$
It is not hard to obtain the explicit formulas
\bea
\tilde{A}_{3}(p,q) = - {q^{2} p\cdot P\over 2 N} \tilde{d}_{m,m,m}(p,q) 
+ {q^{2}\over N} [ \tilde{f}_{3}(p,q) - \tilde{f}_{2}(p,q)] 
+ {p\cdot q \over N} [ \tilde{f}_{3}(p,q) - \tilde{f}_{1}(p,q)] 
\nonumber \\
\tilde{B}_{3}(p,q) = - \tilde{A}_{3}(q,p)
\label{d12}
\eea

The expression
$
\tilde{d}^{\mu}_{2}(p,q)
$
can be obtained from the preceding expression
$
\tilde{d}^{\mu}_{3}(p,q)
$
applying the transformation
\be
p \rightarrow - p,~q \rightarrow P
\label{T23}
\ee
and expression
$
\tilde{d}^{\mu}_{1}(p,q)
$
can be obtained from the expression
$
\tilde{d}^{\mu}_{2}(p,q)
$
applying the transformation
\be
p \rightarrow - q,~q \rightarrow - p.
\label{T12}
\ee

Now we consider the distributions
\be
d^{\mu\nu}_{jk} \equiv {\cal D}_{j}^{\mu}{\cal D}_{k}^{\nu}d_{m,m,m}
\ee
and we have the following generic form of the Fourier transform:
\be
\tilde{d}^{\mu\nu}_{jk}(p,q) = - [p^{\mu}p^{\nu}~\tilde{A}_{jk}(p,q) +
q^{\mu}q^{\nu}~\tilde{B}_{jk}(p,q) + p^{\mu}q^{\nu}~\tilde{C}^{(1)}_{jk}(p,q)
+ q^{\mu}p^{\nu}~\tilde{C}^{(2)}_{jk}(p,q)] + \eta^{\mu\nu}~\tilde{D}_{jk}(p,q)
\label{d21}
\ee
where, as before, the scalar functions
$
A, B, C, D
$
depend only on the Lorentz invariants. 

It is a long but straightforward computation to derive the following
expressions:
\bea
\tilde{A}_{33}(p,q) = {3 q^{2} \over 2 N^{2}} \alpha(p,q) 
+ {1 \over N} \alpha_{2}(p,q)
- {q^{2}\over N} \tilde{f}_{3}(p,q) 
+  {m^{2} q^{2} \over 2N} \tilde{d}_{m,m,m}(p,q)
\nonumber \\
\tilde{B}_{33}(p,q) = {3 p^{2} \over 2 N^{2}} \alpha(p,q) 
+ {1 \over N} \alpha_{1}(p,q)
- {p^{2}\over N} \tilde{f}_{3}(p,q) 
+  {m^{2} p^{2} \over 2N} \tilde{d}_{m,m,m}(p,q) = \tilde{A}_{33}(q,p)
\nonumber \\
\tilde{C}_{33}^{(1)}(p,q) = \tilde{C}_{33}^{(2)}(p,q) 
= - {3 p\cdot q \over 2 N^{2}} \alpha(p,q) 
- {1 \over N} \alpha_{3}(p,q)
+ {p \cdot q\over N} \tilde{f}_{3}(p,q) 
-  {m^{2} p \cdot q \over 2N} \tilde{d}_{m,m,m}(p,q)
\label{d22}
\eea
where
\bea
\alpha_{1}(p,q) = {1\over 4}~(p^{2})^{2}~\tilde{d}_{m,m,m}(p,q)
+ {1 \over 2}~(p^{2} - p \cdot q)~\tilde{f}_{2}(p,q)
- \left(p^{2} - {1\over 2}~p \cdot q\right)~\tilde{f}_{3}(p,q)
\nonumber \\
\alpha_{2}(p,q) = {1\over 4}~(q^{2})^{2}~\tilde{d}_{m,m,m}(p,q)
+ {1 \over 2}~(q^{2} - p \cdot q)~\tilde{f}_{1}(p,q)
- \left(q^{2} - {1\over 2}~p \cdot q\right)~\tilde{f}_{3}(p,q)
\nonumber \\
\alpha_{3}(p,q) = - {1\over 4}~p^{2} q^{2}~\tilde{d}_{m,m,m}(p,q)
- {1 \over 2}~p^{2}~\tilde{f}_{1}(p,q)- {1 \over 2}~q^{2}~\tilde{f}_{2}(p,q) 
\nonumber \\
+ {1\over 2} (p^{2} + q^{2} - p \cdot q)~\tilde{f}_{3}(p,q)
\label{23}
\eea
and
\be
\alpha(p,q) = q^{2}~\alpha_{1}(p,q) + p^{2}~\alpha_{2}(p,q) 
- 2 p \cdot q~\alpha_{3}(p,q).
\label{24}
\ee

The expression
$
\tilde{d}^{\mu}_{22}(p,q)
$
can be obtained from the preceding expression
$
\tilde{d}^{\mu}_{33}(p,q)
$
applying the transformation (\ref{T23}) and expression
$
\tilde{d}^{\mu}_{11}(p,q)
$
can be obtained from the expression
$
\tilde{d}^{\mu}_{22}(p,q)
$
applying the transformation (\ref{T12}). 

In the same way we have
\be
\tilde{D}_{12}(p,q) = - {1 \over 2 N} [q^{2}  \beta_{1}(p,q)
+ p^{2} \beta_{2}(p,q) ] + {p \cdot q \over 2 N} [\beta_{3}(p,q)
+ \beta_{4}(p,q) ] - {1\over 2} \beta_{5}(p,q)
\label{d31}
\ee
and
\bea
\tilde{A}_{12}(p,q) = - {1 \over N} [3 q^{2}  \tilde{D}_{12}(p,q)
+ q^{2} \beta_{5}(p,q) - \beta_{2}(p,q) ]
\nonumber \\
\tilde{B}_{12}(p,q) = - {1 \over N} [3 p^{2}  \tilde{D}_{12}(p,q)
+ p^{2} \beta_{5}(p,q) - \beta_{1}(p,q) ]
\nonumber \\
\tilde{C}_{12}^{(1)}(p,q) = {1 \over N} [3 p \cdot q  \tilde{D}_{12}(p,q)
- \beta_{3}(p,q) + p \cdot q \beta_{5}(p,q) ]
\nonumber \\
\tilde{C}_{12}^{(2)}(p,q) = {1 \over N} [3 p \cdot q  \tilde{D}_{12}(p,q)
- \beta_{4}(p,q) + p \cdot q \beta_{5}(p,q) ].
\label{d32}
\eea

Here we have the notations:
\bea
\beta_{1}(p,q) = - {1\over 4}~p^{2}~(p^{2} + 2 p \cdot q)~\tilde{d}_{m,m,m}(p,q)
- {1\over 2} (p^{2} - p \cdot q)~\tilde{f}_{2}(p,q) 
- {1\over 2} (p\cdot q)~\tilde{f}_{3}(p,q)
\nonumber \\
\beta_{2}(p,q) = - {1\over 4}~q^{2}~(q^{2} + 2 p \cdot q)~\tilde{d}_{m,m,m}(p,q)
- {1\over 2} (q^{2} - p \cdot q)~\tilde{f}_{1}(p,q) 
- {1\over 2} (p\cdot q)~\tilde{f}_{3}(p,q)
\nonumber \\ 
\beta_{3}(p,q) = 
- {1\over 4}~(p^{2} + 2 p \cdot q)~(q^{2} + 2 p \cdot q)~\tilde{d}_{m,m,m}(p,q)
\nonumber \\
- {1\over 2} (p^{2} + 2 p \cdot q)~\tilde{f}_{1}(p,q) 
- {1\over 2} (q^{2} + 2 p \cdot q)~\tilde{f}_{2}(p,q) 
+ {1\over 2} (p^{2} + q^{2} + 3 p\cdot q)~\tilde{f}_{3}(p,q)
\nonumber \\
\beta_{4}(p,q) = 
- {1\over 4}~p^{2}~q^{2}~\tilde{d}_{m,m,m}(p,q)
+ {1\over 2} p^{2}~\tilde{f}_{1}(p,q) 
+ {1\over 2} q^{2}~\tilde{f}_{2}(p,q) 
- {1\over 2} (p^{2} + q^{2} + p\cdot q)~\tilde{f}_{3}(p,q)
\nonumber \\
\beta_{5}(p,q) = 
- {1\over 2}~(p + q)^{2}~\tilde{d}_{m,m,m}(p,q)
- \tilde{f}_{1}(p,q) - \tilde{f}_{2}(p,q) 
+ m^{2}~\tilde{d}_{m,m,m}(p,q)
\nonumber \\
\label{d33}
\eea
The expression
$
\tilde{d}^{\mu}_{13}(p,q)
$
can be obtained from the preceding expression
$
\tilde{d}^{\mu}_{12}(p,q)
$
applying the transformation (\ref{T23}) and expression
$
\tilde{d}^{\mu}_{23}(p,q)
$
can be obtained from the expression
$
\tilde{d}^{\mu}_{13}(p,q)
$
applying the transformation (\ref{T12}).

Using these formulas we can perform the central causal splitting of the
formulas (\ref{shell1}) - (\ref{shell5}).
\begin{thm}
The central splitting of formula (\ref{shell1}) gives for the corresponding
advanced distributions
\be
\eta_{\rho\sigma} (d^{\rho\sigma}_{jj})^{\rm adv} + m^{2}~d^{\rm adv}_{m,m,m}
- 2 f_{j}^{\rm adv} = A~\delta(X)~\delta(Y)
\ee
where
$
A = {i \over 8 (2\pi)^{2}}.
$
\label{s1}
\end{thm}
{\bf Proof:} We work in momentum space and use the formulas (\ref{central1})
and (\ref{central2}). Using formula (\ref{shell1}) one can prove that the
anomaly
\be
\tilde{\cal A}_{j} \equiv 
\eta_{\rho\sigma} (\tilde{d}^{\rho\sigma}_{jj})^{\rm adv} 
+ m^{2}~\tilde{d}^{\rm adv}_{m,m,m} -  2 \tilde{f}_{j}^{\rm adv}
\ee 
is given by the following formula
\be
\tilde{\cal A}_{j}(p,q) = - {i m^{2} \over 2\pi} \int {dt\over t}
\tilde{d}_{m,m,m}(tp, tq).
\ee
The reason of this anomaly is the fact that for the distributions
$
d^{\mu\nu}_{jj}
$
and
$
f_{j}
$
of canonical dimension $0$ we must use the splitting formula (\ref{central2})
and for the distribution
$
d_{m,m,m}
$
of canonical dimension $- 2$ we must use the splitting formula (\ref{central1}).
The integral from the preceding formula has been computed in \cite{Sc1} using
(\ref{d-mmm}) and the result is
\be
a \equiv \int {dt\over t} \tilde{d}_{m,m,m}(tp, tq) 
= - {i \over 8 (2\pi)^{5} m^{2}}.
\label{a}
\ee
Going in the coordinate space we obtain the formula from the statement.
$\qed$

In the same way we have
\begin{thm}
The central splitting of formula (\ref{shell2}) gives for the corresponding
advanced distributions
\bea
\eta_{\rho\sigma} (d^{\mu\rho\sigma}_{3jj})^{\rm adv} 
+ m^{2}~(d^{\mu}_{j})^{\rm adv} - (f^{\mu}_{j3})^{\rm adv} 
= B~(\partial_{1}^{\mu} - \partial_{2}^{\mu})~\delta(X)~\delta(Y)
\nonumber \\
\eta_{\rho\sigma} (d^{\mu\rho\sigma}_{2jj})^{\rm adv} 
+ m^{2}~(d^{\mu}_{j})^{\rm adv} - (f^{\mu}_{j2})^{\rm adv} 
= B~(\partial_{3}^{\mu} - \partial_{1}^{\mu})~\delta(X)~\delta(Y)
\nonumber \\
\eta_{\rho\sigma} (d^{\mu\rho\sigma}_{1jj})^{\rm adv} 
+ m^{2}~(d^{\mu}_{j})^{\rm adv} - (f^{\mu}_{j1})^{\rm adv} 
= B~(\partial_{2}^{\mu} - \partial_{3}^{\mu})~\delta(X)~\delta(Y)
\eea
where
$
B = {1 \over 3}~A.
$
\label{s2}
\end{thm}
{\bf Proof:} As in the preceding formula the anomaly (in the momentum space) is:
\be
\tilde{\cal A}_{jk}^{\mu} \equiv 
\eta_{\rho\sigma} (\tilde{d}^{\mu\rho\sigma}_{jkk})^{\rm adv} 
+ m^{2}~(\tilde{d}^{\mu}_{j})^{\rm adv} - (\tilde{f}_{jk}^{\mu})^{\rm adv}
\ee 
and by the same mechanism as before we have:
\be
\tilde{\cal A}_{jk}^{\mu}(p,q) = - {i m^{2} \over 2\pi} \int {dt\over t^{2}}
\tilde{d}^{\mu}_{j}(tp, tq).
\ee

We must use the formula (\ref{d11}) and we obtain:
\be
\tilde{\cal A}_{jk}^{\mu}(p,q) = - { m^{2} \over 2\pi} 
\left[ p^{\mu}~\int {dt\over t}~\tilde{A}_{j}(tp, tq)
+ q^{\mu}~\int {dt\over t}~\tilde{B}_{j}(tp, tq)\right].
\ee
To compute the two integrals above we must use the formulas (\ref{d12}). For
instance we have:
\bea
\int {dt\over t}~\tilde{A}_{3}(tp, tq)
= - {q^{2} p\cdot P\over 2 N} \int {dt\over t}\tilde{d}_{m,m,m}(tp,tq) 
\nonumber \\
+ {q^{2}\over N} \int {dt\over t^{3}}
[ \tilde{f}_{3}(tp,tq) - \tilde{f}_{2}(tp,tq)] 
+ {p\cdot q \over N} \int {dt\over t^{3}}
[ \tilde{f}_{3}(tp,tq) - \tilde{f}_{1}(tp,tq)]. 
\eea
The first integral has been already computed at the preceding theorem. If we
use the expressions (\ref{f}) then we get
\be
\int {dt\over t^{3}}~\tilde{f}_{1}(tp,tq) = b(p^{2}),\qquad
\int {dt\over t^{3}}~\tilde{f}_{2}(tp,tq) = b(q^{2}),\qquad
\int {dt\over t^{3}}~\tilde{f}_{3}(tp,tq) = b(P^{2})
\ee
where
\be
b(k) \equiv {1\over (2\pi)^{2}}~\int {dt\over t^{3}}~\tilde{d}_{m,m}(tk).
\ee
The preceding integral can be computed using the explicit expression
(\ref{d-mm}) and the result is
\be
b(k) = b~k^{2},\qquad b \equiv - {1 \over 48 (2\pi)^{5} m^{2}}.
\label{b}
\ee
so after some simple substitutions we obtain the formulas from the statement.
$\qed$

We continue the procedure:
\begin{thm}
The central splitting of formula (\ref{shell3}) gives for the corresponding
advanced distributions
\bea
\eta_{\rho\sigma} (d^{\mu\nu\rho\sigma}_{jkll})^{\rm adv} 
+ m^{2}~(d^{\mu}_{jk})^{\rm adv} - (f^{\mu\nu}_{jkl})^{\rm adv} 
= C~\left( a_{jk}^{\mu\nu} 
+ {2 \over 3} C_{l}^{\mu\nu}\right)~\delta(X)~\delta(Y)
\eea
where
$
C = {1 \over 6}~A.
$
Here we have defined the differential operators
\bea
a_{11}^{\mu\nu} \equiv \partial_{2}^{\mu}\partial_{2}^{\nu}
+ \partial_{3}^{\mu}\partial_{3}^{\nu}
- {1\over 2} (\partial_{2}^{\mu}\partial_{3}^{\nu}
+ \partial_{3}^{\mu}\partial_{2}^{\nu})
- {1\over 2}~\eta^{\mu\nu}~(\square_{2} + \square_{3} 
+ \partial_{2}\cdot\partial_{3})
\nonumber \\
a_{12}^{\mu\nu} \equiv - \partial_{1}^{\mu}\partial_{1}^{\nu}
- \partial_{2}^{\mu}\partial_{2}^{\nu}
- {1\over 2} (\partial_{1}^{\mu}\partial_{2}^{\nu}
+ \partial_{2}^{\mu}\partial_{1}^{\nu})
- {1\over 2}~\eta^{\mu\nu}~(\square_{2} + \square_{3} 
+ \partial_{2}\cdot\partial_{3})
\eea
and 
$
a_{22}^{\mu\nu}, a_{33}^{\mu\nu}, a_{23}^{\mu\nu}, a_{31}^{\mu\nu}
$
by circular permutations. The differential operators
$
C_{l}^{\mu\nu}
$
have been defined at theorem \ref{shell}.
\label{s3}
\end{thm}
{\bf Proof:} Formula (\ref{shell3}) can be written as
\bea
{\cal D}_{j}^{\mu}{\cal D}_{k}^{\nu}({\cal K}_{l} + m^{2})d_{m,m,m} =
f^{\mu\nu}_{jkl} + g^{\mu\nu}_{l}
\nonumber \\
g^{\mu\nu}_{l} \equiv - {2 m^{2} \over 3}~C^{\mu\nu}f^{\prime}_{l}
\nonumber 
\eea
and the anomaly is, in momentum space:
\be
\tilde{\cal A}_{jkl}^{\mu\nu} \equiv 
\eta_{\rho\sigma} (\tilde{d}^{\mu\nu\rho\sigma}_{jkll})^{\rm adv} 
+ m^{2}~(\tilde{d}^{\mu\nu}_{jk})^{\rm adv} - (\tilde{f}_{jkl}^{\mu})^{\rm adv}
- (\tilde{g}_{l}^{\mu})^{\rm adv}
\ee 
and by the same mechanism as before we have:
\be
\tilde{\cal A}_{jk}^{\mu\nu}(p,q) = - {i m^{2} \over 2\pi} \int {dt\over t^{3}}
\tilde{d}^{\mu\nu}_{jk}(tp, tq)
+ {i \over 2\pi} \int {dt\over t^{3}} \tilde{g}^{\mu\nu}_{jk}(tp, tq).
\ee
If we use (\ref{d21}) we obtain:
\bea
\tilde{\cal A}_{jk}^{\mu\nu}(p,q) =
{i m^{2}\over 2 \pi}~p^{\mu}p^{\nu}~\int {dt\over t}\tilde{A}_{jk}(tp,tq) 
+ {i m^{2}\over 2 \pi}~q^{\mu}q^{\nu}~\int {dt\over t}\tilde{B}_{jk}(tp,tq) 
\nonumber \\
+ {i m^{2}\over 2 \pi}~p^{\mu}q^{\nu}~
\int {dt\over t}\tilde{C}^{(1)}_{jk}(tp,tq)
+ {i m^{2}\over 2 \pi}~q^{\mu}p^{\nu}~
\int {dt\over t}\tilde{C}^{(2)}_{jk}(tp,tq) 
\nonumber \\
-\eta^{\mu\nu}~{i m^{2}\over 2 \pi}~
\int {dt\over t^{3}}\tilde{D}_{jk}(tp,tq)
+ {i m^{2}\over 3 \pi}~(p_{l}^{\mu} p_{l}^{\nu} - \eta^{\mu\nu} p_{l}^{2})
\int {dt\over t} \tilde{f}_{l}^{\prime}(tp,tq)
\eea
where
$
p_{1} \equiv p, p_{2} \equiv q, p_{3} = - P.
$
If we substitute the formulas for the functions
$
\tilde{A}_{jk}(p,q)
$,
etc. obtained previously then we need beside (\ref{a}), (\ref{b}) a few more
integrals; the first is:
\be
a^{\prime} \equiv \int {dt\over t^{3}} \tilde{d}_{m,m,m}(tp, tq).
\label{a-prime}
\ee
Proceeding as in \cite{Sc1} we obtain
\be
a^{\prime} = {b \over m^{2}}~(p^{2} + q^{2} + p \cdot q).
\label{a-prime1}
\ee
Finally we need
\be
\int {dt\over t} \tilde{f}^{\prime}_{j}(tp, tq) = b.
\label{f-prime}
\ee
Using all these formulas we obtain the result from the statement.
$\qed$

We continue with
\begin{thm}
The central splitting of formula (\ref{shell4}) gives for the corresponding
advanced distributions
\bea
\eta^{\mu\nu}\eta_{\rho\sigma} (d^{\mu\nu\rho\sigma}_{jjkk})^{\rm adv} 
- m^{4}~d_{m,m,m}^{\rm adv} + 2 m^{2} (f_{j}^{\rm adv} + f_{k}^{\rm adv})
= - (A m^{2} + C A_{jk})~\delta(X)~\delta(Y)
\eea
where
\bea
A_{11} \equiv 3 \square_{2} + 3 \square_{3} 
+ 7 \partial_{2}\cdot\partial_{3}
\nonumber \\
A_{12} \equiv \square_{1} + \square_{2} 
- \partial_{1}\cdot\partial_{2}
\eea
and 
$
A_{22}, A_{33}, A_{23}, A_{31}
$
by circular permutations.
\label{s4}
\end{thm}
{\bf Proof:} The anomaly is, in momentum space:
\be
\tilde{\cal A}_{jk} \equiv 
\eta_{\mu\nu}\eta_{\rho\sigma} (\tilde{d}^{\mu\nu\rho\sigma}_{jjkk})^{\rm adv} 
- m^{4}~\tilde{d}_{m,m,m}^{\rm adv} + 2 m^{2} (\tilde{f}_{j}^{\rm adv}
+ \tilde{f}_{k}^{\rm adv})
\ee 
and by the same mechanism as before we have:
\be
\tilde{\cal A}_{jk}(p,q) =  {i m^{4} \over 2\pi} 
\left[\int {dt\over t^{3}} \tilde{d}(tp, tq)
+ \int {dt\over t} \tilde{d}(tp, tq)\right]
- {i m^{2} \over \pi} \int {dt\over t^{3}} [\tilde{f}_{j}(tp, tq)
+ \tilde{f}_{k}(tp, tq)].
\ee
If we use the formulas (\ref{a}), (\ref{b}), (\ref{a-prime1}) and
(\ref{f-prime}) then we obtain the anomaly from the statement.
$\qed$

Finally we have:
\begin{thm}
The central splitting of formula (\ref{shell5}) gives for the corresponding
advanced distributions
\bea
\eta_{\mu\nu}\eta_{\rho\sigma} (d^{\mu\nu\rho\sigma}_{jkll})^{\rm adv} 
- f_{jkl}^{\rm adv} + 2 m^{2} f_{l}^{\rm adv}
= - C D_{jkl}~\delta(X)~\delta(Y)
\eea
where
\bea
D_{111} \equiv 3 \square_{2} + 3 \square_{3} 
+ 7 \partial_{2}\cdot\partial_{3}
\nonumber \\
D_{112} \equiv 3 \square_{2} + \square_{3} 
+ 3 \partial_{2}\cdot\partial_{3}
\nonumber \\
D_{113} \equiv \square_{2} + 3 \square_{3} 
+ 3 \partial_{2}\cdot\partial_{3}
\nonumber \\
D_{231} \equiv 5 \square_{2} + 5 \square_{3} 
+ 9 \partial_{2}\cdot\partial_{3}
\nonumber \\
D_{232} \equiv 5 \square_{2} + 3 \square_{3} 
+ 5 \partial_{2}\cdot\partial_{3}
\nonumber \\
D_{233} \equiv 3 \square_{2} + 5 \square_{3} 
+ 5 \partial_{2}\cdot\partial_{3}
\eea
and the other operators
$
D_{jkl}
$
by cyclic permutations.
\label{s5}
\end{thm}
{\bf Proof:} The anomaly is, in momentum space:
\be
\tilde{\cal A}_{jkl} \equiv 
\eta_{\mu\nu}\eta_{\rho\sigma} (\tilde{d}^{\mu\nu\rho\sigma}_{jkll})^{\rm adv} 
+ m^{2}~\eta_{\mu\nu} (\tilde{d}_{jk}^{\mu\nu})^{\rm adv} 
-\tilde{f}_{jkl}^{\rm adv} + 2 m^{2} \tilde{f}_{l}^{\rm adv}
\ee 
and by the same mechanism as before we have:
\be
\tilde{\cal A}_{jkl}(p,q) =  - {i m^{2} \over 2\pi} 
\eta_{\mu\nu} \int {dt\over t^{3}} \tilde{d}^{\mu\nu}_{jk}(tp, tq)
- {i m^{2} \over \pi} \int {dt\over t^{3}} \tilde{f}_{k}(tp, tq).
\ee
If we use the formulas (\ref{a}), (\ref{b}), (\ref{a-prime1}) and
(\ref{f-prime}) then we obtain the anomaly from the statement.
$\qed$

We point out again that the origin of the anomalies is the fact
that the operation of (central) causal splitting and the operation of taking
the contraction with the Minkowski metric
$
\eta_{\cdot \cdot}
$
do {\bf not} commute. This is the point of the last five theorems.

In the third order of perturbation theory other causal distributions can
appear. These causal distributions are associated to the one-particle reducible
graphs.
\bea
d^{(1)}_{D_{1},D_{2}}(x,y,z) \equiv 
\bar{D}^{F}_{1}(x - y) D_{2}(z - x) - D_{1}(x - y) D^{F}_{2}(z - x)
\nonumber \\
+ D^{(-)}_{1}(x - y) D^{(+)}_{2}(z - x) - D^{(+)}_{1}(x - y) D^{(-)}_{2}(z - x)
]
\nonumber \\
d^{(2)}_{D_{1},D_{2}}(x,y,z) \equiv 
- \bar{D}^{F}_{1}(x - y) D_{2}(y - z) + D_{1}(x - y) D^{F}_{2}(y - z)
\nonumber \\
+ D^{(+)}_{1}(x - y) D^{(-)}_{2}(y - z) - D^{(-)}_{1}(x - y) D^{(+)}_{2}(y - z)
]
\nonumber \\
d^{(3)}_{D_{1},D_{2}}(x,y,z) \equiv 
D^{F}_{1}(z - x) D_{2}(y - z) - D_{1}(z - x) D^{F}_{2}(y - z)
\nonumber \\
+ D^{(-)}_{1}(z - x) D^{(+)}_{2}(y - z) - D^{(+)}_{1}(z - x) D^{(-)}_{2}(y - z)
]
\eea
The causal support properties follow from the alternative formulas
\bea
d^{(1)}_{D_{1},D_{2}}(x,y,z) =
D^{\rm ret}_{1}(x - y) D^{\rm ret}_{2}(z - x) 
- D^{\rm adv}_{1}(x - y) D^{\rm adv}_{2}(z - x)
\nonumber \\
d^{(2)}_{D_{1},D_{2}}(x,y,z) = D^{\rm ret}_{1}(y - x) D^{\rm ret}_{2}(z - y) 
- D^{\rm adv}_{1}(y - x) D^{\rm adv}_{2}(z - y)
\nonumber \\
d^{(3)}_{D_{1},D_{2}}(x,y,z) = D^{\rm ret}_{1}(z - x) D^{\rm ret}_{2}(y - z) 
- D^{\rm adv}_{1}(z - x) D^{\rm adv}_{2}(y - z).
\eea

The order of singularity of these distributions is again
$
\omega = - 2
$.
We can define associated distributions as before if we replace
$
D_{1} \mapsto \partial_{\alpha}D_{1}
$,
etc. 
\bea
{\cal D}^{2}_{\alpha}d^{(1)}_{D_{1},D_{2}} =
d^{(1)}_{D_{1},\partial_{\alpha}D_{2}}, 
\qquad
{\cal D}^{3}_{\alpha}d^{(1)}_{D_{1},D_{2}} =
d^{(1)}_{\partial_{\alpha}D_{1},D_{2}}, 
\nonumber \\
{\cal D}^{1}_{\alpha}d^{(2)}_{D_{1},D_{2}} =
d^{(2)}_{D_{1},\partial_{\alpha}D_{2}}, 
\qquad
{\cal D}^{3}_{\alpha}d^{(2)}_{D_{1},D_{2}} =
d^{(2)}_{\partial_{\alpha}D_{1},D_{2}}, 
\nonumber \\
{\cal D}^{3}_{\alpha}d^{(3)}_{D_{1},D_{2}} =
d^{(3)}_{D_{1},\partial_{\alpha}D_{2}}, 
\qquad
{\cal D}^{2}_{\alpha}d^{(3)}_{D_{1},D_{2}} =
d^{(3)}_{\partial_{\alpha}D_{1},D_{2}}.
\eea
As before we have
\bea
{\partial \over \partial x^{\alpha}}d^{(1)} = 
( {\cal D}^{3}_{\alpha} - {\cal D}^{2}_{\alpha})d^{(1)},
\qquad
{\partial \over \partial y^{\alpha}}d^{(1)} =
- {\cal D}^{3}_{\alpha}d^{(1)}
\qquad
{\partial \over \partial z^{\alpha}}d^{(1)} =
{\cal D}^{2}_{\alpha}d^{(1)}
\nonumber \\
{\partial \over \partial x^{\alpha}}d^{(2)} = 
{\cal D}^{3}_{\alpha}d^{(2)},
\qquad
{\partial \over \partial y^{\alpha}}d^{(2)} =
({\cal D}^{1}_{\alpha} - {\cal D}^{3}_{\alpha})d^{(2)}
\qquad
{\partial \over \partial z^{\alpha}}d^{(2)} =
- {\cal D}^{1}_{\alpha}d^{(2)}
\nonumber \\
{\partial \over \partial x^{\alpha}}d^{(3)} = 
- {\cal D}^{2}_{\alpha}d^{(3)},
\qquad
{\partial \over \partial y^{\alpha}}d^{(3)} =
{\cal D}^{1}_{\alpha}d^{(3)}
\qquad
{\partial \over \partial z^{\alpha}}d^{(3)} =
({\cal D}^{2}_{\alpha} - {\cal D}^{1}_{\alpha})d^{(3)}.
\eea

Now we have relations similar to those from theorem \ref{shell}. First we note
that we have two distinct cases
$
D_{1} = D_{m},~D_{2} = d_{m,.m}
$
and the other way round
$
D_{1} = d_{m,.m},~D_{2} = D_{m}
$
so we define accordingly
\be
d^{(j)}_{m,m,m} = d_{D_{m},d_{m,m}}^{(j)},\qquad 
f^{(j)}_{m,m,m} = d_{d_{m,m},D_{m}}^{(j)}
\ee
and we have:
\begin{thm}
The following relations are true
\be
({\cal K}_{2} + m^{2})d^{(3)} = f_{2},
\label{shell-1PR-1}
\ee
\be
{\cal D}_{1}^{\mu}({\cal K}_{2} + m^{2})d^{(3)} = \partial_{2}^{\mu}f_{2},
\label{shell-1PR-2}
\ee
\be
{\cal D}_{2}^{\mu}({\cal K}_{2} + m^{2})d^{(3)} = - \partial_{1}^{\mu}f_{2},
\label{shell-1PR-3}
\ee
\be
{\cal D}_{1}^{\mu}{\cal D}_{1}^{\nu}({\cal K}_{3} + m^{2})d^{(3)} 
= \partial_{2}^{\mu}\partial_{2}^{\nu}f_{2},
\label{shell-1PR-4}
\ee
\be
{\cal D}_{2}^{\mu}{\cal D}_{2}^{\nu}({\cal K}_{2} + m^{2})d^{(3)} 
= \partial_{1}^{\mu}\partial_{1}^{\nu}f_{3},
\label{shell-1PR-5}
\ee
\be
{\cal D}_{1}^{\mu}{\cal D}_{2}^{\nu}({\cal K}_{2} + m^{2})d^{(3)} 
= - \partial_{2}^{\mu}\partial_{1}^{\nu}f_{3}
\label{shell-1PR-6}
\ee
and similar relations for the other five distributions of this type.
These relations can be causality split without anomalies.
\label{shell-1PR}
\end{thm}

{\bf Proof:} We can proceed as in the proceeding theorems but there is a simple
way, namely to notice that we take the causal split to be
\bea
d^{(1)adv}_{D_{1},D_{2}}(x,y,z) = D^{\rm ret}_{1}(x - y) D^{\rm ret}_{2}(z - x),
\quad
d^{(1)ret}_{D_{1},D_{2}}(x,y,z) = D^{\rm adv}_{1}(x - y) D^{\rm adv}_{2}(z - x)
\nonumber \\
d^{(2)adv}_{D_{1},D_{2}}(x,y,z) = D^{\rm ret}_{1}(y - x) D^{\rm ret}_{2}(z - y),
\quad 
d^{(2)ret}_{D_{1},D_{2}}(x,y,z) =  D^{\rm adv}_{1}(y - x) D^{\rm adv}_{2}(z - y)
\nonumber \\
d^{(3)adv}_{D_{1},D_{2}}(x,y,z) = D^{\rm ret}_{1}(z - x) D^{\rm ret}_{2}(y - z),
\quad
d^{(3)ret}_{D_{1},D_{2}}(x,y,z) = D^{\rm adv}_{1}(z - x) D^{\rm adv}_{2}(y - z)
\eea
and similar relations for the associated distributions
$
{\cal D}^{2}_{\alpha}d^{(1)}_{D_{1},D_{2}}
$,
etc. 
$\qed$

\newpage
\section{Anomalies in the Third Order of the Perturbation Theory}

We remind that by $s$ we have denoted the cohomology operator of the causal
formalism (see the Introduction). We want to compute the one-loop contributions
from
$
(sD)^{IJK};
$
there will be a piece coming from the triangle graphs (here the
distribution
$
d_{D_{1},D_{2},D_{3}}
$
will appear) and another from 1-particle reducible graphs (where the
distributions
$
d^{(j)}_{D_{1},D_{2}}
$
play the central role). The computations are very long and perhaps the easiest
way is to use the off-shell formalism developed in a previous paper
\cite{caciulata4}. We first consider the contributions even with respect
to parity. We need some definitions:
\be
f^{(0)}_{[abc]} = f_{eap} f_{ebq} f_{cpq},\quad
\quad
f^{(1)}_{abc} = f^{\prime}_{pae} f^{\prime}_{qbe} f^{\prime}_{pqc},\quad
\quad
f^{(3)}_{[abc]} = f^{\prime}_{epa} f^{\prime}_{eqb} f^{\prime}_{pqc},\quad
f^{(4)}_{[abc]} = - i~Tr([ t_{a}^{\epsilon},t_{b}^{\epsilon}]t_{c}^{\epsilon})
\ee
and 
\bea
t^{(1)}_{a\epsilon} = \sum_{b} g_{ab}~t^{\epsilon}_{b},\quad
t^{(2)}_{a\epsilon} = \sum_{b}
t^{\epsilon}_{b}t^{\epsilon}_{a}t^{\epsilon}_{b},\quad
t^{(3)}_{a\epsilon} = \sum_{b} 
s^{- \epsilon}_{b}t^{\epsilon}_{a}s^{\epsilon}_{b}.
\eea

Then we have for instance in the top ghost number sector for triangle
graphs:
\bea
(sD)^{[\mu][\nu]\emptyset}_{\rm even, triangle}(x,y,z)
\nonumber \\
= [ - (2 f^{(0)}_{abc} + f^{(3)}_{abc})
({\cal D}_{3}^{\mu}{\cal D}_{3}^{\nu}{\cal K}_{1} 
+ {\cal D}_{2}^{\mu}{\cal D}_{2}^{\nu}{\cal K}_{1} 
+ {\cal D}_{2}^{\mu}{\cal D}_{3}^{\nu}{\cal K}_{1} 
+ {\cal D}_{3}^{\mu}{\cal D}_{2}^{\nu}{\cal K}_{1} 
- \eta^{\mu\nu}~{\cal K}_{1}~{\cal K}_{3})
\nonumber \\
+ 2~f^{(4)}_{abc}
({\cal D}_{2}^{\mu}{\cal D}_{3}^{\nu}{\cal K}_{1} 
+ {\cal D}_{3}^{\mu}{\cal D}_{2}^{\nu}{\cal K}_{1} 
- \eta^{\mu\nu}~{\cal D}_{2}\cdot{\cal D}_{3}~{\cal K}_{1})]d_{m,m,m}(x,y,z)~
u_{a}(x) u_{b}(y) u_{c}(z) 
\nonumber \\
- (x \leftrightarrow y, \mu \leftrightarrow \nu)
+ \cdots
\label{sd2tr}
\eea
where by
$
\cdots
$
we mean super-renormalizable terms. We also have
\be
(sD)^{\emptyset\emptyset[\mu\nu]}_{\rm even, triangle}(x,y,z) = 0.
\ee
If we consider the 1-particle reducible graphs then we have
\bea
(sD)^{[\mu][\nu]\emptyset}_{\rm even, 1PR}(x,y,z)
\nonumber \\
= {1\over 3}~(2 f^{(0)}_{abc} + f^{(3)}_{abc} + 2 f^{(4)}_{abc} )
({\cal D}_{2}^{\mu}{\cal D}_{2}^{\nu} 
- \eta^{\mu\nu}~{\cal K}_{2})~{\cal K}_{1}f^{(3)}(x,y,z)~
u_{a}(y) u_{b}(y) u_{c}(z) 
\nonumber \\
+ ({\cal D}_{3}^{\mu}{\cal D}_{3}^{\nu}
- \eta^{\mu\nu}~{\cal K}_{3})~{\cal K}_{1}f^{(2)}(x,y,z)~
u_{a}(x) u_{b}(z) u_{c}(z) 
\nonumber \\
- (x \leftrightarrow y, \mu \leftrightarrow \nu)
\label{sd21pr}
+ \cdots
\eea
and
\be
(sD)^{\emptyset\emptyset[\mu\nu]}_{\rm even, 1PR}(x,y,z) = 0.
\ee
It is a consistency check to use theorems \ref{shell} and \ref{shell-1PR} to
prove that the sum of the two expressions (\ref{sd2tr}) and (\ref{sd21pr}) cancel
on-shell. 

Now we define the advanced (retarded, Feynman) operators substituting in
$
D^{IJK}_{\rm even, triangle}
$
the distribution
$
d_{m,m,m}
$
(and associated ones) by the corresponding distributions
$
d_{m,m,m}^{\rm adv}, d_{m,m,m}^{\rm ret}
$
and
$
d_{m,m,m}^{F},
$
etc. We do the similar substitutions in
$
D^{IJK}_{\rm even, triangle}
$
and we obtain anomalies because of theorems \ref{s1} - \ref{s5}. After some
computations we obtain from the preceding formulas the anomaly:
\bea
{\cal A}^{[\mu][\nu]\emptyset}_{\rm even}(x,y,z)
= - {2 C\over 3} (2 f^{(0)}_{abc} + f^{(3)}_{abc} + 8 f^{(4)}_{abc} )~
[\partial_{1}^{\mu}\partial_{1}^{\nu} - \partial_{2}^{\mu}\partial_{2}^{\nu}
- \eta^{\mu\nu}(\square_{1} - \square_{2})]~\delta(X)~\delta(Y)
\nonumber \\
u_{a}(x) u_{b}(z) u_{c}(z)
\eea
and
\be
{\cal A}^{\emptyset\emptyset[\mu\nu]}_{\rm even}(x,y,z) = 0.
\ee
Proceeding in the same way we obtain
\bea
{\cal A}^{\emptyset\emptyset[\mu]}_{\rm even}(x,y,z)
= {2 C\over 3} (2 f^{(0)}_{abc} + f^{(3)}_{abc} + 8 f^{(4)}_{abc} )~
[\partial_{1}^{\mu}\partial_{1}^{\nu} - \partial_{2}^{\mu}\partial_{2}^{\nu}
- \eta^{\mu\nu}(\square_{1} - \square_{2})]~\delta(X)~\delta(Y)
\nonumber \\
v_{a\nu}(x) u_{b}(z) u_{c}(z) + ( x \leftrightarrow y)
\eea
and
\bea
{\cal A}^{\emptyset\emptyset\emptyset}_{\rm even}(x,y,z)
= - {2 C\over 3} (2 f^{(0)}_{abc} + f^{(3)}_{abc} + 8 f^{(4)}_{abc} )~
[\partial_{1}^{\mu}\partial_{1}^{\nu} - \partial_{2}^{\mu}\partial_{2}^{\nu}
- \eta^{\mu\nu}(\square_{1} - \square_{2})]~\delta(X)~\delta(Y)
\nonumber \\
v_{a\nu}(x) v_{b\nu}(z) u_{c}(z) 
\nonumber \\
- 3 B f^{(1)}_{abc} (\partial_{1}^{\mu} + \partial_{2}^{\mu})\delta(X)~\delta(Y)
[ \partial_{\mu}\Phi_{a}(x) \Phi_{b}(y) u_{c}(z)
- \Phi_{a}(x) \partial_{\mu}\Phi_{b}(y) u_{c}(z)]
\nonumber \\
+ 2 B f^{(2)}_{abc} [ (\partial_{1}^{\mu} + 2 \partial_{2}^{\mu})
\delta(X)~\delta(Y) \partial_{\mu}\Phi_{a}(x) \Phi_{b}(y) u_{c}(z)
\nonumber \\
- (2 \partial_{1}^{\mu} + \partial_{2}^{\mu})
\delta(X)~\delta(Y) \Phi_{a}(x) \partial_{\mu}\Phi_{b}(y) u_{c}(z)]
\nonumber \\
- C f^{(5)}_{abc} (\square_{1} - \square_{2})~\delta(X)~\delta(Y)
\Phi_{a}(x) \Phi_{b}(y) u_{c}(z)
\nonumber \\
- 3 B (\partial_{1}^{\mu} + \partial_{2}^{\mu})\delta(X)~\delta(Y)
u_{a}(z) \left[\bar{\psi}(x) 
\left( {1\over 2}t^{(1)}_{a\epsilon} + 2t^{(2)}_{a\epsilon} +
t^{(3)}_{a\epsilon}\right)
\psi(y) + ( x \leftrightarrow y)\right]
\nonumber \\
+ ( x \leftrightarrow z) + ( y \leftrightarrow z) 
\eea
where we do not give the complicated expression of
$
f^{(5)}_{abc}
$
because in fact it can be proved that the preceding anomaly is a coboundary.
\begin{thm}
The following formula is verified:
\be
{\cal A}^{IJK}_{\rm even}(x,y,z) = (sB)^{IJK}(x,y,z).
\label{anomaly}
\ee
We can take
\be
B^{\emptyset\emptyset\emptyset}(x,y,z) = 
a_{abc}~(\partial_{1}^{\mu} - \partial_{2}^{\mu})\delta(X)~\delta(Y)~
v_{a\nu}(x)~v_{b}^{\nu}(y)~v_{c\mu}(z) 
+ 3 b_{abc}\delta(X)~\delta(Y)~v_{a\mu}(x)~v_{b\nu}(y)~F_{c}^{\mu\nu}(z) 
\ee
where we must have
\be
a_{abc} + b_{abc} = {2 i\over 3}~C~(2 f^{(0)}_{abc} + f^{(3)}_{abc} + 8 f^{(4)}_{abc}).
\ee
\end{thm}
{\bf Proof:} We must start with the generic form of the cocycle $B$. We first consider the pure Yang-Mils sector. 
We have two types of terms: one of the form
$
\partial\delta(X)~\delta(Y)~W(x,y,z)
$
with $W$ of canonical dimension $3$ and
$
\delta(X)~\delta(Y)~W(x,y,z)
$
with $W$ of canonical dimension $4$. The first sector has the following expression in top ghost dimension:
\bea
B_{1}^{[\mu][\nu][\rho]}(x,y,z) = 
[ a^{(1)}_{abc}~(\eta^{\mu\rho} \partial_{1}^{\nu} - \eta^{\nu\rho} \partial_{2}^{\mu})
\nonumber \\
+ a^{(2)}_{abc}~(\eta^{\mu\rho} \partial_{2}^{\nu} - \eta^{\nu\rho} \partial_{1}^{\mu})
+ a^{(3)}_{abc}~\eta^{\mu\nu}~( \partial_{1}^{\rho} - \partial_{2}^{\rho})]~\delta(X)~\delta(Y)
u_{a}(x) u_{b}(y) u_{c}(z)
\nonumber \\
B_{1}^{[\mu\rho][\nu]\emptyset}(x,y,z) = 
[ b^{(1)}_{abc}~(\eta^{\mu\nu} \partial_{1}^{\rho} - \eta^{\nu\rho} \partial_{1}^{\mu})
+ b^{(2)}_{abc}~(\eta^{\mu\nu} \partial_{2}^{\rho} - \eta^{\nu\rho} \partial_{2}^{\mu})]
\nonumber \\
~\delta(X)~\delta(Y)~u_{a}(x) u_{b}(y) u_{c}(z).
\eea
If we substitute in (\ref{anomaly}) we get after some computations some constraints on the free parameters above:
\be
a^{(1)}_{abc} = 2 a_{abc}, \quad a^{(2)}_{abc} = a_{abc}, \quad a^{(3)}_{abc} = - a_{abc}
\ee
and
\be
a_{abc} + b^{(1)}_{abc} = {2i\over 3} C (2 f^{(0)}_{abc} + f^{(3)}_{abc} + 8 f^{(4)}_{abc}).
\ee
If we consider the expression
$
{\cal A}^{IJK}_{\rm even}(x,y,z) - (sB_{1})^{IJK}(x,y,z)
$
we find out terms of the form
$
\partial\delta(X)~\delta(Y)~F(x) u(y) u(z)
$
which can be eliminated with considering new coboundaries of the form
$
\delta(X)~\delta(Y)~F(x) u(y) u(z)
$
and in the end we obtain the assertion from the statement in the pure Yang-Mills sector. The scalar and the Dirac sectors 
can be treated in the same way and they do not produce new constraints.
$\qed$

We remark the fact that the redefinition of the chronological products which must be done to eliminate the anomalies
\be
T^{IJK}(x,y,z) \rightarrow T^{IJK}(x,y,z) + B^{IJK}(x,y,z)
\ee
does produce physical effects in the null ghost sector
$
I = J = K = \emptyset.
$

Now we consider the anomaly in the sector odd with respect to parity. After some computations we obtain 
\bea
(sD)^{[\mu][\nu]\emptyset}_{\rm odd}(x,y,z) =
- 2 i~\varepsilon^{\mu\nu\rho\sigma}~
({\cal D}_{1\rho}{\cal D}_{3\sigma}{\cal K}_{2} - {\cal D}_{2\rho}{\cal D}_{3\sigma}{\cal K}_{1})
d_{m,m,m}(x,y,z)
\nonumber\\
A_{abc}~u_{a}(x) u_{b}(z) u_{c}(z)
\nonumber \\
(sD)^{\emptyset\emptyset[\mu\nu]}_{\rm odd}(x,y,z) = 0
\nonumber \\
(sD)^{\emptyset\emptyset[\mu]}_{\rm odd}(x,y,z) =
- 2 i~\varepsilon^{\mu\nu\rho\sigma}
({\cal D}_{1\rho}{\cal D}_{2\sigma}{\cal K}_{3} - {\cal D}_{1\rho}{\cal D}_{3\sigma}{\cal K}_{2})
d_{m,m,m}(x,y,z)
\nonumber \\
A_{abc}~u_{a}(x) v_{b\nu}(z) u_{c}(z) + (x \leftrightarrow y)
\nonumber \\
(sD)^{\emptyset\emptyset\emptyset}_{\rm odd}(x,y,z) =
- 2 i~\varepsilon^{\mu\nu\rho\sigma}
({\cal D}_{1\rho}{\cal D}_{3\sigma}{\cal K}_{2} - {\cal D}_{2\rho}{\cal D}_{3\sigma}{\cal K}_{1})
d_{m,m,m}(x,y,z)
\nonumber \\
~A_{abc}~u_{a}(x) v_{b\nu}(z) u_{c}(z) 
+ (x \leftrightarrow z) + (y \leftrightarrow z)
\eea
where 
\be
A_{abc} \equiv \sum_{\epsilon}~\epsilon~Tr(\{t^{\epsilon}_{a}, t^{\epsilon}_{b}\} t^{\epsilon}_{c})
\ee
is a symmetric tensor; there are no contributions in the $1$-particle reducible sector.

If we use the theorem \ref{s3} the nontrivial anomalies are:
\bea
{\cal A}^{[\mu][\nu]\emptyset}_{\rm odd}(x,y,z)
= - {1 \over 3(2 \pi)^{2}}~\varepsilon^{\mu\nu\rho\sigma} \partial_{1\rho}\partial_{2\sigma}
~\delta(X)~\delta(Y)~A_{abc}~u_{a}(x) u_{b}(z) u_{c}(z)
\nonumber \\
{\cal A}^{\emptyset\emptyset[\mu]}_{\rm odd}(x,y,z)
= - {1 \over 3(2 \pi)^{2}}~\varepsilon^{\mu\nu\rho\sigma} \partial_{1\rho}\partial_{2\sigma}
~\delta(X)~\delta(Y)~A_{abc}~u_{a}(x) v_{b\nu}(z) u_{c}(z) + ( x \leftrightarrow y)
\nonumber \\
{\cal A}^{\emptyset\emptyset\emptyset}_{\rm odd}(x,y,z)
= - {1 \over 3(2 \pi)^{2}}~\varepsilon^{\mu\nu\rho\sigma} \partial_{1\rho}\partial_{2\sigma}
~\delta(X)~\delta(Y)~A_{abc}~v_{a\mu}(x) v_{b\nu}(z) u_{c}(z) 
\nonumber \\
+  ( x \leftrightarrow z) +  ( y \leftrightarrow z)
\eea
 If we write the generic form of a possible coboundary we can easily find out that the
the relation
\be
{\cal A}^{IJK}_{\rm odd}(x,y,z) = (sB)^{IJK}(x,y,z)
\ee
is possible {\it iff} 
\be
B^{IJK}(x,y,z) = 0 \Longleftrightarrow {\cal A}^{IJK}_{\rm odd}(x,y,z) = 0 \Longleftrightarrow A_{abc} = 0
\ee
i.e. the axial anomaly should be null.

We have investigated the anomalies of the standard model of maximal canonical dimension 
$
\omega = 5
$
in the third order of the perturbation theory. Anomalies of lower canonical dimension must be investigated 
separately. 
\section{Conclusions}
We have proved that the one-loop contributions to the chronological products can 
produce anomalies only in orders 
$
n = 2, 3, 4
$
of the perturbation theory. We proved that if we can eliminate the anomalies in these orders, then we 
will not have one-loop anomalies for higher orders of the perturbation
theory. The key point was to prove that some identities involving distributions
can be extended without anomalies.

Next, we have determined the generic form of the one-loop anomalies of maximal canonical dimension in the orders 2 and 3
of the perturbation theory. The origin of these anomalies is the causal splitting of some relations
where contractions with the Minkowski metric do appear. We still have to analyze the order $4$.  Also in order
$
n = 2, 3
$
we still have to analyze anomalies of lower dimension. Cohomology methods might work in this case.
The generalization of the preceding analysis to multi-loop contributions in not obvious and it is a
subject of further investigation.

\end{document}